\documentclass[preprint,amsmath,amssymb,superscriptaddress,showpacs,pre]{revtex4-1}

\usepackage{graphicx}% Include figure files
\usepackage{dcolumn}% Align table columns on decimal point
\usepackage{bm}% bold math
\usepackage{color}
\usepackage{epstopdf} 
\usepackage{amsmath,amsfonts,amssymb,amsthm}
\usepackage{algorithm}
\usepackage{algorithmic}
\usepackage{braket}
\usepackage[sans]{dsfont}
\usepackage{mathbbol}
\usepackage{bbm}
\usepackage{subfigure}

\usepackage{lineno}
\usepackage{nicefrac}       % compact symbols for 1/2, etc.
\usepackage{microtype}      % microtypography

\let\originalleft\left
\let\originalright\right
\renewcommand{\left}{\mathopen{}\mathclose\bgroup\originalleft}
\renewcommand{\right}{\aftergroup\egroup\originalright}

\newcommand{\ie}{\emph{i.e.}}

\def\<{\langle}
\def\>{\rangle}
\def\vx{\vec x}

\DeclareMathOperator{\Tr}{Tr}

\newcommand{\Lam}{\bm{\Lambda}}
\newcommand{\Sig}{\bm{\Sigma}}

\newcommand{\curlynormalpar}[1]{\exp\left\{-\frac{1}{2}\left( #1 \right)\right\}}
\newcommand{\curlynormal}[1]{\exp\left\{-\frac{1}{2} #1 \right\}}
\newcommand{\iSig}{\bm{\Sigma^{-1}}}
\newcommand{\iC}{\bm{C}^{-1}}
\newcommand{\av}[1]{\left\langle #1 \right\rangle}
\newcommand{\vsa}{\vec{s}_a}
\newcommand{\vuka}{\vec{u}_{ka}}

\newcommand{\sref}[1]{S\ref{#1}}

\begin{document}

\title{Global multivariate model learning from hierarchically correlated data}
% \date{}

\author{Edwin Rodr\'{i}guez Horta} 
\affiliation{Sorbonne Universit\'{e}, CNRS, Institut de Biologie
  Paris-Seine, Laboratoire de Biologie Computationnelle et
  Quantitative -- LCQB, Paris, France}
\affiliation{Group of Complex Systems and Statistical Physics, Department of Theoretical Physics, University of Havana, Havana, Cuba}
\author{Alejandro Lage} 
\affiliation{Group of Complex Systems and Statistical Physics, Department of Theoretical Physics, University of Havana, Havana, Cuba}
\author{Martin Weigt} 
\affiliation{Sorbonne Universit\'{e}, CNRS, Institut de Biologie
  Paris-Seine, Laboratoire de Biologie Computationnelle et
  Quantitative -- LCQB, Paris, France}

\author{Pierre Barrat-Charlaix} 
\email{Correspondence to: Pierre Barrat-Charlaix, \bf{pierre.barrat@unibas.ch}}
\affiliation{Biozentrum, Universit\"at Basel, Basel, Switzerland}

\begin{abstract}
Inverse statistical physics aims at inferring models compatible with a set of empirical averages estimated from a high-dimensional dataset of independently distributed equilibrium configurations of a given system. However, in several applications such as biology, data result from stochastic evolutionary processes, and configurations are related through a hierarchical structure, typically represented by a tree, and therefore not independent. In turn, empirical averages of observables superpose intrinsic signal related to the equilibrium distribution of the studied system and spurious historical (or phylogenetic) signal resulting from the structure underlying the data-generating process. The naive application of inverse statistical physics techniques therefore leads to systematic biases and an effective reduction of the sample size. To advance on the currently open task of extracting intrinsic signals from correlated data, we study a system described by a multivariate Ornstein-Uhlenbeck process defined on a finite tree. Using a Bayesian framework, we can disentangle covariances in the data corresponding to their multivariate Gaussian equilibrium distribution from those resulting from the historical correlations. Our approach leads to a clear gain in accuracy in the inferred equilibrium distribution, which corresponds to an effective two- to fourfold increase in sample size.
\end{abstract}

\maketitle

\section{Introduction}
\label{sec:int}

With the emergence of large, high-dimensional datasets for complex systems across disciplines, methods of {\it inverse statistical physics} have seen rapidly growing interest during the last years \cite{Inverse_problem_Berg}. In the most standard setting, the data provide observational samples of the ``microscopic''degrees of freedom of the system under study -- this can be biological sequences \cite{levy_potts_2017,cocco_inverse_2018}, firing patterns of neurons \cite{schneidman2006weak,roudi2009ising}, individuals in animal groups \cite{bialek2012statistical,cavagna2018physics}, stock markets \cite{bury2013market,borysov2015us} etc.
Within a static modeling approach, frequently based on the maximum-entropy principle \cite{jaynes1957information}, data $\vec x$ are assumed to be generated independently from some unknown probability distribution $P(\vec x)$. This distribution describes the underlying interaction patterns between the observed degrees of freedom, and has to be learned from data to unveil the rules governing the system. 
In more rare cases where data correspond to observed time series, theoretical and algorithmic development is much less advanced than for independent static data \cite{Inverse_problem_Berg}.

One of the biggest application areas of inverse statistical mechanics is the modeling of biological processes.
These applications are fuelled by the large amount of available data resulting from the impressive progress in experimental techniques in biology. 
This is especially visible in the case of biological sequences,  with databases now harboring a vast amount of high-quality DNA or protein sequences \cite{sayers_genbank_2019,uniprotconsortium_uniprot_2018}.
A common idea in this context is that it is possible to use characteristics of genes or organisms related by a common ancestry -- called \emph{homologous} -- to construct models of the selection acting on them. 
A successful example in this regard is the representation of protein sequences by probabilistic models in the so-called DCA method \cite{levy_potts_2017,cocco_inverse_2018}. 
The prototypical datasets in this context are multiple-sequence alignments (MSA), with lines being a so-called homologous, {\em i.e.} evolutionarily related sequences, and columns specific positions deriving from some common ancestral position \cite{durbin1998biological}.  
The MSA contains at least two kinds complementary information:
\begin{itemize}
    \item {\it Phylogenetic information:} the distances between sequences carry information about the evolutionary time since their common ancestor. Using phylogeny inference methods \cite{felsenstein_phylogenies_1988,felsenstein2004inferring} we may reconstruct the evolutionary history of our dataset, represented by a phylogenetic tree.  
    \item {\it Co-evolutionary information:} positions in a sequence typically do not evolve independently, but rather in a correlated way. This co-evolution carries important information about the selection forces acting on evolving entities. This fact has been extensively studied in the case of protein sequences, and used to predict structure, mutational landscapes or networks of interacting proteins \cite{levy_potts_2017,cocco_inverse_2018}.  
\end{itemize}
These two types of information are contained in two complementary features of the data: phylogenetic inference is based on the comparative analysis of different sequences, while co-evolutionary information is contained in the correlation of different columns of the MSA. 
Modeling approaches using one type of information typically neglect the other one: inference of phylogenies generally assumes that all positions in a sequence evolve independently, while co-evolutionary models of proteins assume that sequences in the MSA are independently distributed. 
This choice is motivated by the fact that taking the two types of correlations into account, \emph{i.e.} through time with phylogeny and accross trait values for co-evolution, results in very hard inference procedures, cf.~\cite{obermayer2014inverse,rodriguez2019toward}. 
However, this can lead to biases in the model parameters: it has for instance been shown that phylogenetic relations between protein sequences induce non-trivial correlations that are not related to protein function \cite{qin_power_2018,horta2020phylogenetic}. 

In this work, we consider the case of the inverse problem for high-dimensional data showing hierarchical correlations due to a branching generating process.
Our motivation for this purely methodological study comes from the modeling of protein sequences discussed above, but the underlying problem is much more general. 
Instead of sequences of discrete characters, like amino acids or nucleotides, we may consider continuous phenotypic traits. 
The branching process is not necessarily the phylogeny of species, but it may be the genealogy of populations of the same species, or other branching processes like epidemics spreading or geographic migration.

To address this problem, we use a simple and very general model for the temporal evolution of correlated variables: a historically well-known way to represent such processes is to use Ornstein-Uhlenbeck dynamics (OU), which models configurations as Gaussian vectors evolving in a quadratic potential that represents selection forces \cite{uhlenbeck_theory_1930, felsenstein_phylogenies_1988, hansen_stabilizing_1997}.
OU processes are commonly used in the field of phylogenetic comparative methods (PCM) \cite{bartoszek_phylogenetic_2012,mitov_fast_2020}. 
This modeling approach is \emph{a priori} limited to continuous traits, but could potentially be used for protein sequences combined with a continuous-variable approximation, that has successfully been used in the past \cite{jones_psicov_2012,barton_large_2014,baldassi_fast_2014}. 
In this context, the equilibrium distribution reached by the OU process represents the probability distribution given by the DCA method, which can be used to predict non-trivial structural contacts in the protein fold, effects of amino-acid mutations or even designing novel functional sequences \cite{morcos_direct-coupling_2011, figliuzzi_coevolutionary_2016, russ_evolution-based_2020}.

In this work, we are interested in constructing an inference method for parameters of an OU process from data correlated through a tree.
Our approach is purely methodological, and the data can represent any set of continuous phenotypic traits, \emph{e.g.} from different organisms, with the tree indicating the phylogenetic relations between data points. 
Inferred parameters then represent the selection forces without biasing effects from the phylogeny. 
The manuscript is divided as follows: we first review in section \ref{sec:ornstein_uhlenbeck_dynamics} the main characteristics of the multivariate OU process. 
We then describe the setting of the inference problem that we want to solve in section \ref{sub:statement_of_the_problem}, propose a solution in sections \ref{sub:Calculation_of_the_likelihood} and \ref{sub:maximization_of_the_likelihood}.
Finally, we present results obtained on simulated data in section \ref{sec:Results}, with the context of pairwise models of protein sequences in mind.

\section{The multivariate Ornstein-Uhlenbeck process}
\label{sec:ornstein_uhlenbeck_dynamics}

We consider a system characterized by $L$ continuous degrees of freedom and whose state is fully described by an $L$-dimensional vector $\vx\in\mathbb{R}^L$. These degrees of freedom can be continuous phenotypic traits of some living organism, or the sequence of a gene or a protein if a continuous approximation is made. 
At equilibrium, $\vx$ is assumed to be normally distributed,
\begin{equation}
	P_{eq}(\vx) = \frac{1}{Z(\bm{J})}\curlynormal{\vx^T\bm{J}\vx}\ ,
	\label{eq:eq_distribution}
\end{equation}
where $\bm{J}$ is the symmetric, positive definite \emph{coupling matrix} and $Z(\bm{J}) = \sqrt{(2\pi)^L/\det \bm{J}}$ is the normalization constant; the means of all components of $\vx$ are set to zero without loss of generality. 
We are interested in inferring the coupling matrix from a given amount of observed states $\vx$ of the system.
If these observations were independent from each other, due to the simple Gaussian form of Eq.~\eqref{eq:eq_distribution}, $\bm{J}$ would simply be equal to the inverse of the empirical \emph{covariance matrix} of the data, written $\bm{C}=\bm{J}^{-1}$. 

However, we consider the case where observations are not independent. 
On the contrary, they result from a dynamical process taking place during a finite amount of time, and different data-points are therefore correlated to each other.
This dynamical process is described below.

We suppose that the considered system evolves according to the following Langevin equation 
\begin{equation}
	\gamma^{-1}\frac{\text{d}\vx}{\text{d}t} = - \bm{J}\vx + \vec{\xi}(t).
	\label{eq:langevin}
\end{equation}
Here, $\vec{\xi}(t)$ is a vector of uncorrelated white noise, and $\gamma^{-1}$ is the characteristic timescale governing the dynamics. 
In short, Eq.~\eqref{eq:langevin} states that the system described by $\vx$ undergoes Brownian motion in a quadratic energy landscape characterized by the coupling matrix $\bm{J}$. 

We are not interested in $\vx$ directly, but rather in its probability distribution $P(\vx\vert\,\vx_0,\Delta t)$, \ie~in the probability to find the system in state $\vx$ knowing it was in state $\vx_0$ some time $\Delta t$ in the past. 
The Fokker-Planck equation corresponding to Eq.~\eqref{eq:langevin} is straightforward to write, 
\begin{equation}
	\gamma^{-1} \partial_t P = \left(-\sum_{a,b=1}^L\frac{\partial}{\partial x_a} J_{ab}x_b + \sum_{a=1}^L\frac{\partial^2}{\partial x_a^2}\right) P, 
	\label{eq:FP}
\end{equation}
where the parenthesized expression on the right hand side is understood as an operator acting on $P$. 
The solution to Eq.~\eqref{eq:FP} is a multivariate normal distribution~\cite{singh2017multiOU}:
\begin{equation}
	P(\vx|\, \vx_0, \Delta t) = \left[(2\pi)^N\det\Sig\right]^{-1/2}
	\curlynormal{(\vx-\vec{\mu})^T\iSig(\vx- \vec{\mu})},
	\label{eq:OUpropagator}
\end{equation}
where we introduce the matrices $\Sig$ and $\Lam$ as well as the vector $\vec{\mu}$ as 
\begin{equation}
	\Lam = e^{-\gamma\bm{J}}, \qquad \vec{\mu} = \Lam^{\Delta t} \vx_0, \qquad \Sig = \bm{J}^{-1}(\mathbb{1} - \Lam^{2 \Delta t}).
	\label{eq:def_lambda}
\end{equation}
Eqs.~\eqref{eq:OUpropagator} and~\eqref{eq:def_lambda} define a multivariate \emph{Ornstein-Uhlenbeck} (OU) process. 

Note that since matrix $\Lam$ is an exponential of $\bm{J}$, it is symmetric, has strictly positive eigenvalues and commutes with $\bm{J}$. 
We also underline that $\Sig$ and $\vec{\mu}$ depend on $\Delta t$, although this dependence is not explicitly written in our notation to make it less heavy. 
By taking $\gamma\Delta t \gg 1$ and using the fact that $\bm{J}$ has strictly positive eigenvalues, one immediately recovers Eq.~\eqref{eq:eq_distribution}, meaning that the OU process converges to the desired equilibrium distribution.

We can compute the joint distribution of two configurations $\vx_1$ and $\vx_2$ separated by a time $\Delta t$ by multiplying Eqs.~\eqref{eq:eq_distribution} and~\eqref{eq:OUpropagator}, 
\begin{eqnarray}
    P(\vx_1, \vx_2\vert\,\Delta t) &= & P(\vx_1 |\, \vx_2,\Delta t) \times P_{eq}(\vx_2)
    \nonumber\\
    &\propto& \curlynormalpar{\vx_1^T\iSig\vx_1 + \vx_2^T\iSig\vx_2 - 2\vx_1^T\Lam^{\Delta t}\iSig\vx_2}. 
	\label{eq:OUjoint}	
\end{eqnarray}
This equation illustrates the \emph{time reversibility} of the OU process. 
Indeed, the distribution is symmetric in $\vx_1$ or $\vx_2$ and does not depend on which configuration came first.

Equation \eqref{eq:OUjoint} allows for computing the joint covariance of the correlated equilibrium configurations $\vx_1$ and $\vx_2$. 
The probability distribution in Eq.~\eqref{eq:OUjoint} is normal with an inverse covariance matrix defined by blocks: $\Sig$ on the diagonal and $-\Lam^{\Delta t}\Sig$ off-diagonal. 
By inverting this block matrix, given that $\Lam$ and $\Sig$ commute and are invertible, one obtains the following covariance: 
\begin{equation}
	\langle \vx_1\vx_2^T \rangle_{\Delta t} = \Lam^{\Delta t}\bm{J}^{-1} = \Lam^{\Delta t}\bm{C}. 
	\label{eq:pairwisecov}
\end{equation}

Eq.~\eqref{eq:pairwisecov} allows us to readily distinguish two regimes. 
Let us call $\rho_a$ the eigenvalues of $\bm{J}$. 
The eigenvalues of $\Lam^{\Delta t}\bm{C}$ are then equal to $\rho_a^{-1}e^{-\gamma\rho_a\Delta t}$. 
Since all $\rho_a$ are positive, the eigenvalues of $\Lam^{\Delta t}\bm{C}$ vanish exponentially over time. The slowest timescale of exponential decay is set by $\tau_c^{-1} = \gamma\rho_{min}$, with $\rho_{min}$ being the smallest eigenvalue of $\bm{J}$. 
Thus, for $\Delta t / \tau_c \gg 1$, $\vx_1$ and $\vx_2$ are uncorrelated. 
If this is verified for all pairs of observations $\vx_i$ and $\vx_j$, the regime is that of \emph{uncorrelated} data -- the inference of $\bm{J}$ can simply be performed by inverting the empirical covariance matrix extracted from the data. 
Inversely, for $\Delta t / \tau_c \ll 1$, $\vx_1$ and $\vx_2$ are highly correlated, defining a \emph{strongly correlated} regime. 
It should be noted that for $\Delta t = 0$, the joint correlation matrix of $\vx_1$ and $\vx_2$ becomes non invertible, and Eq.~\eqref{eq:pairwisecov} becomes irrelevant. Actually, $\vx_1$ and $\vx_2$ coincide at that point, \ie~we have $P(\vx_1,\vx_2|\,\Delta t=0) = P_{eq}(\vx_1) \times \delta(\vx_1 - \vx_2)$ using the $L$-dimensional Dirac distribution.

\section{Methods}
\label{sec:methods}

\subsection{Statement of the problem}
\label{sub:statement_of_the_problem}

\begin{figure}[!htb]
	\includegraphics[width=0.8\textwidth]{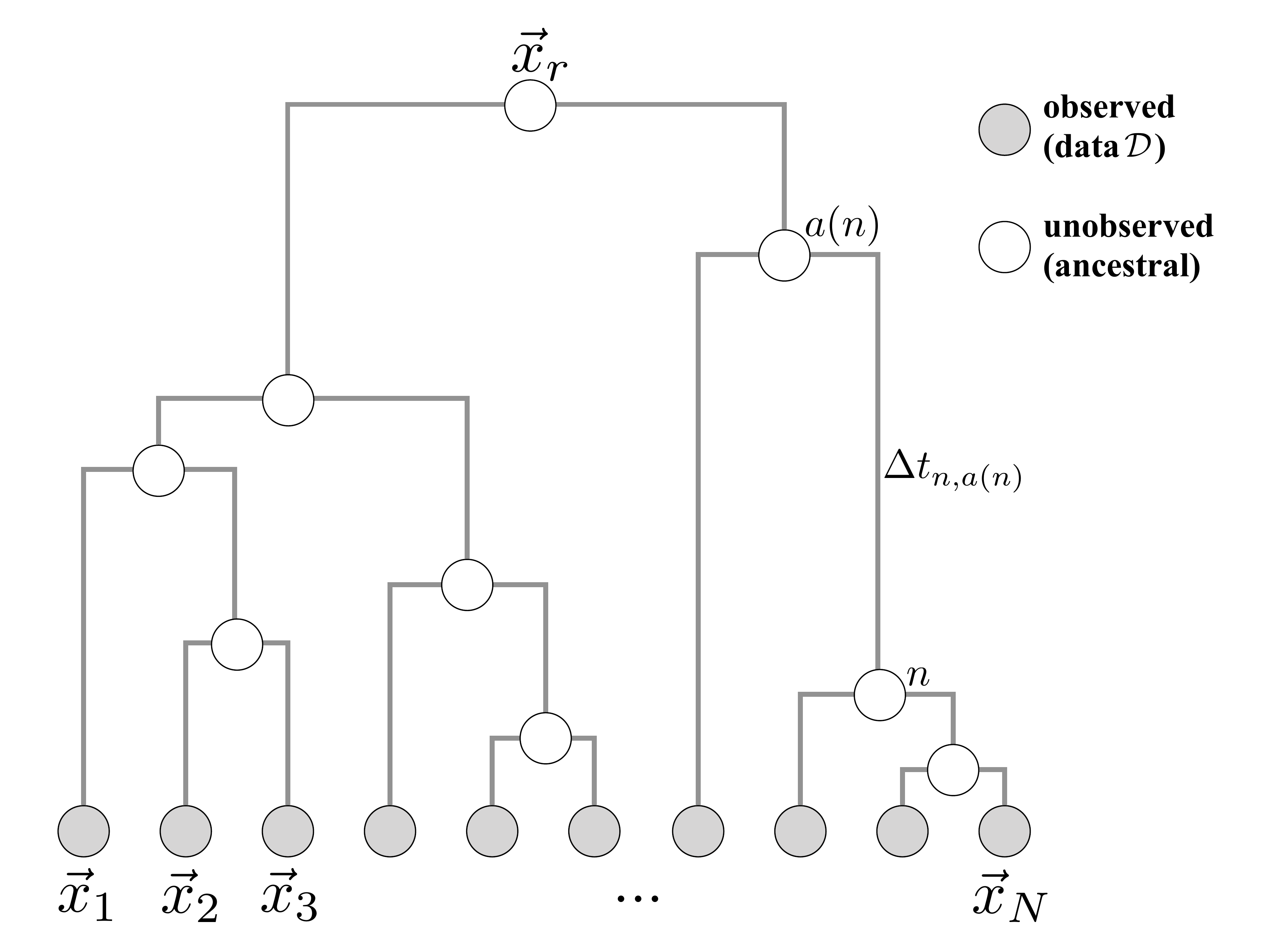}
	\caption{Schematic representation of a tree $\mathcal{T}$ underlying the data generating process. The process starts at the root node $r$ with a configuration $\vx_r$ sampled from $P_{eq}(\vx_r)$. The dynamics consist in independent realizations of the OU process on all branches from ancestral nodes $a(n)$ to child nodes $n$ over times corresponding to the branch length $\Delta t_{n,a(n)}$, initialized in the ancestral configuration $\vx_{a(n)}$. The observable data only consist of configurations of the leaf nodes (grey circles in the figure), while configurations of ancestral nodes remain unknown. There are no restrictions on the topology of tree $\mathcal{T}$ and the length of the branches.}
	\label{fig:sample_tree}
\end{figure}

The problem discussed here is the inference of the probability distribution describing samples that are hierarchically correlated by a tree, cf.~Fig.~\ref{fig:sample_tree}. 
Formally, we assume that the data consists of $N$ real-valued vectors of length $L$, denoted $\{\vx_i\}\in\mathbb{R}^L$ with $i=1,...,N$. 
Taken individually, we assume that the $\vx_i$ are distributed according to Eq.~\eqref{eq:eq_distribution}, \ie~according to a multivariate Gaussian of zero mean and covariance $\bm C$.
By construction, the equilibrium covariance between any pair of elements of a given vector $\vx=(x^1,...,x^L)$ is given by the inverse of the coupling matrix: $\langle x^a x^b\rangle-\langle x^a\rangle\langle x^b\rangle = \bm C_{ab} = (\bm J^{-1})_{ab}$ for all $a,b\in\{1,...,L\}$. 
This implies that inferring the coupling matrix defining the probability distribution amounts to finding the \emph{equilibrium} covariance matrix $\mathbf{C}$.  

However, this covariance cannot be directly measured as we consider observations that are not independently distributed.
Instead, the set of measured configurations $\{\vx_i\}_{i=1,...,N}$ is the result of an Ornstein-Uhlenbeck (OU) process taking place on a tree $\mathcal{T}$, as is illustrated in Fig.~\ref{fig:sample_tree}:
\begin{itemize}
    \item The process starts at the root node $r$ with a state vector $\vx_r$ drawn from the equilibrium distribution $P_{eq}$.
    \item On each branch $(n,a(n))$ of length $\Delta t_{n,a(n)}$ connecting node $n$ with its ancestral node $a(n)$, the dynamics follow Eq.~\eqref{eq:langevin}, starting from initial condition $\vx_{a(n)}$, and running for time $\Delta t_{n,a(n)}$. In other words, given the state $\vx_{a(n)}$ of the ancestral node, $\vx_n$ is sampled from $P(\vx_n |\, \vx_{a(n)}, \Delta t_{n,a(n)})$, see Eq.~\eqref{eq:OUpropagator}
    \item As a consequence, OU processes on branches stemming from a common ancestral node evolve independently, but from an identical initial condition.
    \item Observed data vectors correspond to the states of the leaves of the tree at the end of this process. The states of the internal nodes are not part of the observed data and remain unknown. 
\end{itemize}
This process is thought to represent the evolution of biological traits along a phylogenetic tree, with the leaf nodes corresponding to traits observed in today's species. Note that due to the reversible nature of our OU process, the joint probability of any pair of leaf configurations $\vx_i$ and $\vx_j$, with $i,j\in\{1,...,N\}$, is given by  $P(\vx_i,\vx_j|\,\Delta t_{ij})$ (Eq.~\eqref{eq:OUjoint}), with $\Delta t_{ij}$ denoting the total branch length of the path connecting $i$ and $j$ in the tree.

The OU process is characterized by the quadratic potential $\bm J = \bm C^{-1}$ and the rate $\gamma$. 
Hence, the joint statistics of the leaf configurations $\{\vx_i\}_{i=1,...,N}$ (\ie~the data) is fully determined by $\bm C$, $\gamma$, and the tree $\mathcal{T}$. 
The aim of this work is to derive a method for inferring the most likely values of $\bm C$ and $\gamma$ given the knowledge of the data $\mathcal{D}=\{\vx_i\}_{i=1,...,N}$ and the underlying tree $\mathcal{T}$. We consider here that both the topology and the branch lengths of $\mathcal{T}$ are known. 

This problem shows two notable extreme cases: 
The first one is the case where the typical branch length of the tree is short compared to the timescales of the OU process.
As a consequence, leaf configurations are close to identical to the root, \ie~$\vx_i\simeq\vx_r$, and the inference of $\bm C$ becomes impossible. 
The second one is the opposite case where the typical branch length of the tree is long compared to the longest timescale of the OU process $\tau_c$.
In this case, the configuration of a child node is close to independent from that of its ancestor, and leaf configurations can be considered as independent samples from the equilibrium distribution $P_{eq}$.
$\bm C$ can then be readily estimated by computing the empirical covariance matrix. 
We are interested here in the intermediate regime where substantial tree-mediated correlations between data
make it impossible to simply estimate $\bm C$ with the empirical covariance, but the depth of the tree introduces enough variability in the data for one to hope of reconstructing the energy potential $\bm{J}$.

We adopt a Bayesian inference approach by writing the probability of a given set of parameters $\{\bm C, \gamma\}$ given the data $\{\mathcal{D},\mathcal{T}\}$ using Bayes' equation
\begin{equation}
	P(\bm C, \gamma\vert\mathcal{D},\mathcal{T}) \propto P(\mathcal{D}\vert\bm C, \gamma,\mathcal{T})\cdot P(\bm C, \gamma),
	\label{eq:bayesian_likelihood}
\end{equation}
with the proportionality constant not depending on the parameters $\{\bm C, \gamma\}$. Here, $P(\bm C, \gamma)$ can be any arbitrarily chosen prior distribution. The difficulty in Eq.~\eqref{eq:bayesian_likelihood} lies in the estimation of the likelihood $P(\mathcal{D}\vert\bm C, \gamma, \mathcal{T})$,
\ie~of the joint probability of the datapoints $\mathcal{D}=\{\vx_i\}_{i=1,...,N}$ for an OU process given by its parameters $\{\bm C, \gamma\}$ and the tree $\mathcal{T}$.  We detail the computation of this probability in the following section. 

\subsection{Calculation of the likelihood} % (fold)
\label{sub:Calculation_of_the_likelihood}
 
The joint distribution of two configurations $\vx_1$ and $\vx_2$ separated by time $\Delta t$ is given by Eq.~\eqref{eq:OUjoint} and corresponds to a joint normal distribution. 
This means that the vector $\vec{X}=[\vx_1, \vx_2]$, \emph{i.e.} the concatenation of vectors $\vx_1$ and $\vx_2$, follows a normal distribution with zero mean and variance described above in Eqs.~\eqref{eq:def_lambda}.
Of importance here is that this property of the OU process can be extended to the joint distribution of any subset of nodes in a tree. 
In other words, if we now define $\vec{X}=[\vx_1,\ldots,\vx_N]$ to be the concatenation of all configurations in our dataset $\mathcal D$, we can write the distribution of $\vec{X}$ as
\begin{equation}
	P(\vec{X}\vert\,\bm{C},\gamma,\mathcal{T}) = \left((2\pi)^{LN}\det\mathbb{G}\right)^{-\frac{1}{2}}\curlynormal{\vec{X}^T\mathbb{G}^{-1}\vec{X}},
	\label{eq:multiOU_jointall}
\end{equation}
where $\mathbb{G}$ is the \emph{joint covariance matrix} and depends on the tree as well as on $\bm{C}$ and $\gamma$.

The joint covariance matrix is a matrix of dimension $(L\cdot N)\times(L\cdot N)$, built by $N\times N$ blocks of size $L\times L$ with entries
\begin{equation}
	\mathbb{G}_{ij}(a,b) = \av{x_i^a x_j^b} - \av{x_i^a}\av{x_j^b},\ \ \ \ i,j\in\{1,...,N\}; a,b\in\{1,...,L\},
	\label{eq:jointcov_1}
\end{equation}
where the (zero) marginals $\av{x_i^a}$ and $\av{x_j^b}$ are explicitly written for clarity.
Each block $\mathbb{G}_{ij}$ is describing the connected correlations between two data vectors $\vx_i$ and $\vx_j$, which are separated by time $\Delta t_{ij}$, resulting as the sum of all branch lengths of the path connecting $i$ and $j$ on tree $\mathcal{T}$. Because the OU process is time reversible, we can directly apply Eq.~\eqref{eq:pairwisecov} and give all blocks of $\mathbb{G}$ in closed form, 
\begin{equation}
	\mathbb{G}_{ij} =
	\begin{cases}
		\bm C & \text{if $i=j$}\\
		\Lam^{\Delta t_{ij}}{\bm C} & \text{otherwise}  ,
	\end{cases}
	\label{eq:jointcov_2}
\end{equation}
using the (currently unknown) covariance matrix $\bm C$ of a single equilibrium vector $\vx$. We remind here that $\Lam = e^{-\gamma\iC}$ depends only on $\gamma$ and $\bm C$, and commutes with $\bm C$. As a direct consequence, all blocks $\mathbb{G}_{ij}$ commute with each other and with $\bm C$.

Eq.~\eqref{eq:multiOU_jointall} allows us to compute the log-likelihood of the data $\vec{X}$ as a function of $\vec{X}$ itself and of the joint covariance matrix. Indeed, taking its logarithm immediately gives
\begin{equation}
	\mathcal{L}_{\mathcal{D}}(\mathbb{G}) = -\frac 1 2 \log \det  \mathbb{G}  -\frac 1 2 \vec{X}^T\mathbb{G}^{-1} \vec{X} + \text{const}\ ,
	\label{eq:likelihood_2}
\end{equation}
but this expression is impractical for any numerical evaluation due to the large dimension of $\mathbb{G}$. However, the particular block structure of $\mathbb{G}$ described in Eq.~\eqref{eq:jointcov_2} allows us to simplify the expression. To do so, we first introduce the eigenvalues and eigenvectors 
$\left\{\rho_a, \vsa\right\}$ of $\iC$, where the index $a$ runs from $1$ to $L$ and vectors $\vsa$ are of dimension $L$. By definition, we have $\rho_a>0$ for all $a$.  Using now Eq.~\eqref{eq:jointcov_2}, we immediately see that the vectors $\vsa$ are also eigenvectors of the individual blocks $\mathbb{G}_{ij}$ with eigenvalues $z(\rho_a, \Delta t_{ij})$ where we introduced
\begin{equation}
	z(\rho_a, \Delta t_{ij}) = \rho_a^{-1}e^{-\gamma\rho_a\Delta t_{ij}} \ .
	\label{eq:z}
\end{equation}
By convention, $\Delta t_{ii} = 0$ and the diagonal blocks are thus included via $z(\rho_a, \Delta t_{ii}) = \rho_a^{-1}$.

As the next step, we introduce $N\times N$-dimensional matrices ${\bm G}^a, a=1,...,L,$ with elements 
\begin{equation}
	{\bm G}^a_{ij} = z(\rho_a, \Delta t_{ij})\ ,\ \ \ \ 1\leq i,j\leq N\ .
	\label{eq:subG_def}
\end{equation}
In other words, for a given index $1\leq a\leq L$, ${\bm G}^a$ is the matrix built by replacing all blocks of $\mathbb{G}$ by their respective $a$th eigenvalue.
Matrices ${\bm G}^a$ are symmetric and have their own eigenmodes, that we denote by $\left\{\lambda_{ka}, \vuka\right\}_{k=1,...,N}$.

To obtain the eigenmodes of the joint covariance matrix $\mathbb{G}$ as a function of the $\vsa$ and $\vuka$, we construct the direct product of vectors $\vsa$ and $\vuka$, defining vectors $\vec{S}_{ka}$ of dimension $L\times N$:
\begin{equation}
	\begin{split}
		\vec{S}_{ka} &= \vuka \otimes \vsa \\
		&= [u_{ka}^1\cdot\vsa, \ldots, u_{ka}^N\cdot\vsa].
		% &= [u_{ka}(1)s_a(1), \ldots, u_{ka}(1)s_a(L); \ldots; u_{ka}(N)s_a(1), \ldots, u_{ka}(N)s_a(L)]
	\end{split}
	\label{eq:Sdef}
\end{equation}
The $i$th block vector of $\vec{S}_{ka}$ will thus be written as $\vec{S}_{ka}^i = u_{ka}^i\cdot\vsa$. 
We can now show that $\vec{S}_{ka}$ are eigenvectors of matrix $\mathbb{G}$ by considering the $i$th block vector of the product $\mathbb{G}\cdot\vec{S}_{ka}$:
\begin{equation}
	\begin{split}
		\left(\mathbb{G}\cdot\vec{S}_{ka}\right)^i &= \sum_{j=1}^N \mathbb{G}_{ij} u_{ka}^j\cdot\vsa \\
		&= \sum_{j=1}^N z(\rho_a, \Delta t_{ij}) u_{ka}^j\cdot \vsa \\
		&= (\mathbf{G}^a\cdot\vuka)^i \cdot \vsa\\
		&= \lambda_{ka}(u_{ka}^i\cdot\vsa) \\
		&= \lambda_{ka}\vec{S}_{ka}^i \ .
	\end{split}
\end{equation}
We have first used the fact that $\vsa$ is an eigenvector of $\mathbb{G}_{ij}$, then the definition of ${\bm G}^a$, and finally the fact that $\vuka$ is an eigenvector of ${\bm G}^a$. 
This demonstrates that the eigenmodes of $\mathbb{G}$ are $\left\{ \lambda_{ka}, \vec{S}_{ka} \right\}$ with $1\leq k\leq N$ and $1\leq a \leq L$. 
Since $\mathbb{G}$ is the covariance matrix of a Gaussian distribution, we conclude the $\lambda_{ka}$ to be strictly positive.
Interestingly, the definition of $\vec{S}_{ka}$ as a direct product between eigenvectors $\vsa$ of the energy potential and eigenvectors $\vuka$ reflecting the correlation structure mediated by the tree illustrates how these two types of information are entangled in the covariance matrix of the data.

Note that this decomposition of the eigenvectors leads to a drastic decrease in computational complexity for diagonalizing $\mathbb{G}$ (at given $\bm C$, $\gamma$ and $\cal T$), and in consequence also for calculating the likelihood according to Eq.~\eqref{eq:likelihood_2}, which depends on the inverse covariance matrix $\mathbb{G}^{-1}$. Matrix $\mathbb{G}$ has linear dimension $LN$, so the numerical diagonalization or inversion takes time ${\cal O}( (L N)^3)$. This is hardly achievable for systems of realistic length $L$ of the state vector, and sufficient number $N$ of data points for model learning. Following the above description, we need to first diagonlize $\bm C^{-1}$ (or equivalently $\bm C$), which requires time of ${\cal O}(L^3)$, followed by inversion of the $L$ matrices ${\bm G}^a$, each one having linear dimension $N$. The total time complexity therefore results in ${\cal O}(L^3) + {\cal O}(L\cdot N^3)$, and the calculation can be easily achieved even on a standard PC. This observation is essential for inference, since we need to redo this calculation for many realizations of $\bm C$ and $\gamma$, in order to find the ones maximizing the likelihood given the data $\cal D$ and the tree $\cal T$. As is shown in section \sref{sub:homogeneous_and_fully_balanced_tree}, this calculation simplifies even more when considering a fully balanced and homogeneous tree. In this case, the matrices $\bm G^a$ commute and can be diagonalized simultaneously and analytically for any value of $\rho^a$.

For the case of arbitrary trees, Eq.~\eqref{eq:likelihood_2} can now be rewritten using the eigen-decomposition of $\mathbb{G}$: 
\begin{equation}
	\begin{split}
		\mathcal{L}_{\mathcal{D}}(\mathbb{G}) &= -\frac 1 2 \sum_{k=1}^N\sum_{a=1}^L \log\lambda_{ka} - \frac 1 2 \sum_{k=1}^N\sum_{a=1}^L \lambda_{ka}^{-1}(\vec{X}\cdot \vec{S}_{ka})^2\\
		&= -\frac 1 2 \sum_{k,a} \left(\log\lambda_{ka} + \lambda_{ka}^{-1}	\left(\sum_{i=1}^N u_{ka}^i\vx_i\cdot\vsa\right)^2 \right).
	\end{split}
	\label{eq:likelihood_3}
\end{equation}
Eq.~\eqref{eq:likelihood_3} expresses the likelihood as a function of $\vuka$, $\lambda_{ka}$ (resulting from the tree $\cal T$ and given $\rho^a$) and $\vsa$ (resulting from $\bm C$). 
However, the definition of $\bm G^a$ in Eq.~\eqref{eq:subG_def} makes clear that its eigenmodes $\{\lambda_{ka}, \vuka \}$ depend only of the eigenvalues $\rho_a$ of $\iC$, on $\gamma$, as well as of the structure of the tree through the quantities $\Delta t_{ij}$, although this dependence cannot be analytically expressed in a simple manner. 
This means that the likelihood in equation~\eqref{eq:likelihood_3} is in fact a function of $\{\rho_a, \vsa\}$, \emph{i.e.} the eigenmodes of $\iC$, of the time scale parameter $\gamma$ and of the pairwise distances on the tree $\Delta t_{ij}$.

% subsection derivation_of_the_likelihood (end) 

\subsection{Maximizing the likelihood} % (fold)
\label{sub:maximization_of_the_likelihood}

As stated at the beginning of this section, our main task is to find the equilibrium covariance matrix $\bm C$ that maximizes the likelihood of the data. 
We also need to find the optimal time scale $\gamma$. 
In Eq.~(\eqref{eq:likelihood_3}), the likelihood is expressed as a function of $\gamma$ and $\{\rho_a, \vsa\}$, \ie~the eigenvalues and eigenvectors of $\iC$, either directly or through the quantities $\{\lambda_{ka}, \vuka \}$. 
We know attempt to maximize the likelihood with respect to the eigenmodes $\{\rho_a, \vsa\}$ and to the time scale $\gamma$. 

In order to perform this optimization, we need to compute the gradient of the likelihood with repsect to the eigenvectors $\{\vsa\}$. 
Since $\iC$ is a symmetric matrix, its eigenvectors form an orthogonal basis of the vector-space of dimension $L$ and their components cannot be changed independently.  
One possible parametrization for the $\{\vsa\}$ consists in using $L(L-1)/2$ scalar \emph{Eulerian angles} $\{\theta_{\alpha\beta}\}$ with $1\leq \alpha < \beta \leq L$~\cite{Raffenetti1970GEA, Shepard_param_OM}. 
With the $L$ eigenvalues $\rho_a$, this results in $L(L+1)/2$ independent values that fully parametrize the $L(L+1)/2$ values of $\iC$. 
A second possibility, that we have found faster in practice, is to express the matrix of the $\{\vsa\}$ as the exponential of a skew-symetric matrix with $L(L-1)/2$ independent values, see section \ref{sub:parametrizations_of_eigenvectors} of the appendix. 
However, this parametrization does not allow a simple analytical expression of the gradient of the likelihood, and we use it along with automatic differentiation \cite{Zygote.jl-2018}. 
For this reason, we use the Eulerian angles below to express the gradient of the likelihood. 

As a first step, we need to compute the gradient of the likelihood $\mathcal{L}_{\mathcal{D}}(\mathbb{G})$ with respect to all parameters $\{\rho_a, \theta_{\alpha\beta}\}$ and $\gamma$. 
To make explicit the dependences of eigenvalues and eigenvectors of the matrices $\bm G^a$ on these parameters, we introduce the notation $\vec{u}_k(\rho_a,\gamma) = \vuka$ and $\lambda_k(\rho_a,\gamma) = \lambda_{ka}$. 
Note that from the definition of $\bm G^a$ in Eq.~\eqref{eq:subG_def}, its eigenvalues and vectors depend only on the eigenvalues of $\iC$ and not on its eigenvectors. 
In the same way, we will now write $\bm G(\rho_a,\gamma)$ instead of $\bm G^a$. 

The gradient of the likelihood is obtained by differentiating Eq.~\eqref{eq:likelihood_3} with respect to the parameters of interest. 
This gives us  three equations: 
% dL / d rho
\begin{equation}
	\begin{split}
		\frac{\partial \mathcal{L}}{\partial\rho_a} &= -\frac 1 2 \sum_{k=1}^N \left\{ \frac{\partial\lambda_k}{\partial\rho_a}\lambda_k^{-1} - \frac{\partial\lambda_k}{\partial\rho_a} \lambda_k^{-2}\left(\sum_{i=1}^N u_{k}^i\vx_i\cdot\vsa\right)^2 \right. \\
		&+ \left. 2\lambda_k^{-1} \left(\sum_{i=1}^N u_k^i\vx_i\cdot\vsa\right) \left(\sum_{i=1}^N \frac{\partial u_k^i}{\partial\rho_a} \vx_i\cdot\vsa\right) \right\}, 
	\end{split}
	\label{eq:gradlikelihood_1}
\end{equation}
% dL / d theta
\begin{equation}
	\frac{\partial \mathcal{L}}{\partial\theta_{\alpha\beta}} = \sum_{k=1}^N \lambda_k^{-1} \left(\sum_{i=1}^N u_k^i\vx_i\cdot\vsa\right) \left(\sum_{i=1}^N u_k^i \vx_i\cdot\frac{\partial \vsa}{\partial\theta_{\alpha\beta}}\right),
	\label{eq:gradlikelihood_2}
\end{equation}
and
% dL / d gamma
\begin{equation}
	\begin{split}
		\frac{\partial \mathcal{L}}{\partial\gamma} &= -\frac 1 2 \sum_{k=1}^N \left\{ \frac{\partial\lambda_k}{\partial\gamma}\lambda_k^{-1} - \frac{\partial\lambda_k}{\partial\gamma} \lambda_k^{-2}\left(\sum_{i=1}^N u_{k}^i\vx_i\cdot\vsa\right)^2 \right. \\
		&+ \left. 2\lambda_k^{-1} \left(\sum_{i=1}^N u_k^i\vx_i\cdot\vsa\right) \left(\sum_{i=1}^N \frac{\partial u_k^i}{\partial\gamma} \vx_i\cdot\vsa\right) \right\}, 
	\end{split}
	\label{eq:gradlikelihood_3}
\end{equation}
The derivatives of $\vec{u}_k(\rho,\gamma)$ and $\lambda_k(\rho,\gamma)$ with respect to $\rho$ can then be computed using the following equations~\cite{matrix_cook_book}: 
\begin{equation}
	\frac{\partial\lambda_i(\rho,\gamma)}{\partial\rho} = \vec{u}_k(\rho,\gamma)^T\frac{\partial\bm{G}(\rho,\gamma)}{\partial\rho}\vec{u}_k(\rho,\gamma)
	\label{eq:grad_eigval}
\end{equation}
and
\begin{equation}
	\frac{\partial \vec{u}_k(\rho,\gamma)}{\partial\rho} = \sum_{l\neq k} \left( \vec{u}_k(\rho,\gamma)^T\frac{\partial\bm{G}(\rho,\gamma)}{\partial\rho}\vec{u}_l(\rho,\gamma) \right) \left(\lambda_k(\rho,\gamma) - \lambda_l(\rho,\gamma)\right)^{-1} \vec{u}_l(\rho,\gamma).
	\label{eq:grad_eigvec}
\end{equation}
Equivalent equations can be written for their derivatives with respect to $\gamma$. 

The computation of the gradient of $\mathcal{L}$ for a given set of parameters $\{\rho_a, \theta_{\alpha\beta}\}$ then goes as follows. 
For each eigenvalue $\rho_a$, we compute and diagonalize matrix $\bm{G}(\rho_a)$ to obtain its eigenmodes $\vec{u}_k(\rho_a)$ and $\lambda_k(\rho_a)$. 
Using equations~\eqref{eq:grad_eigval} and ~\eqref{eq:grad_eigvec} and their equivalent form for $\gamma$, we also numerically compute their derivatives with respect to $\rho_a$ and $\gamma$. 
This gives us all the quantities to estimate the gradient of $\mathcal{L}$ with respect to $\rho_a$ using equation~\eqref{eq:gradlikelihood_1}.

 The optimization is performed by a quasi-Newton method \cite{NLopt}. Details are presented in section \ref{sub:optimization_scheme} of the apprendix.

% subsection maximization_of_the_likelihood (end)

\section{Results}
\label{sec:Results}
% \subsection{Choosing an appropriate time scale interval}

In order to evaluate our inference procedure, we generate artificial data corresponding to the process described in section \ref{sub:statement_of_the_problem}. 
We first build a  balanced binary tree $\mathcal{T}$ with $2^9=512$ leaves. 
The length of each branch of $\mathcal{T}$ is chosen from a uniform distribution in the interval $[0,1]$. 
We also sample positive semi-definite coupling matrix $\bm{J}$ of size $L\times L$ with $L=4$ or $L=10$, with entries normally distributed with mean $\mu_J=0.8$ and $\sigma_J=0.2$. 

In the case of statistical models of protein sequence, a major achievement is the ability of pairwise models to predict contacts in the three-dimensional structure of the protein from an inferred coupling matrix.
In order to replicate this setting and to perform interaction prediction, we randomly set to $0$ off-diagonal elements of $J$ with probability $0.7$, resulting in a sparsified coupling matrix of approximate density $0.3$.  
Zero elements of $J$ correspond to variables that do not interact, in analogy to non-contacts in the case of an application to protein sequences.

In order to investigate the different regimes of tree-induced correlation, we vary the parameter $\gamma$ around a reference timescale $\gamma_d$ defined as follows: 
\begin{equation}
    \gamma_d = \frac{1}{\Delta t_{av}\rho_{min}} 
\end{equation}
where $\Delta t_{av}$ is the average branch length separating two leaves of $\mathcal{T}$. 
For $\gamma \gg \gamma_d$, leaf configurations are on average well decorrelated, whereas for $\gamma \ll \gamma_d$ all leaves will be strongly correlated. 
By simulating data using different $\gamma$ in the range $[10^{-2},2]\cdot\gamma_d$, we investigate all relevant temporal regimes.  
For each value of $\gamma$, we then sample  configurations of leaves of $\mathcal{T}$ using the process described in section \ref{sub:generating_artificial_data} of the supplementary material. 
To avoid statistical noise when assessing the quality of our inference, we repeat the sampling of leaf configurations 100 times for each value of $\gamma$.

\begin{figure}[!htb]
	\centering
	\includegraphics[keepaspectratio=true,width=0.5\textwidth]{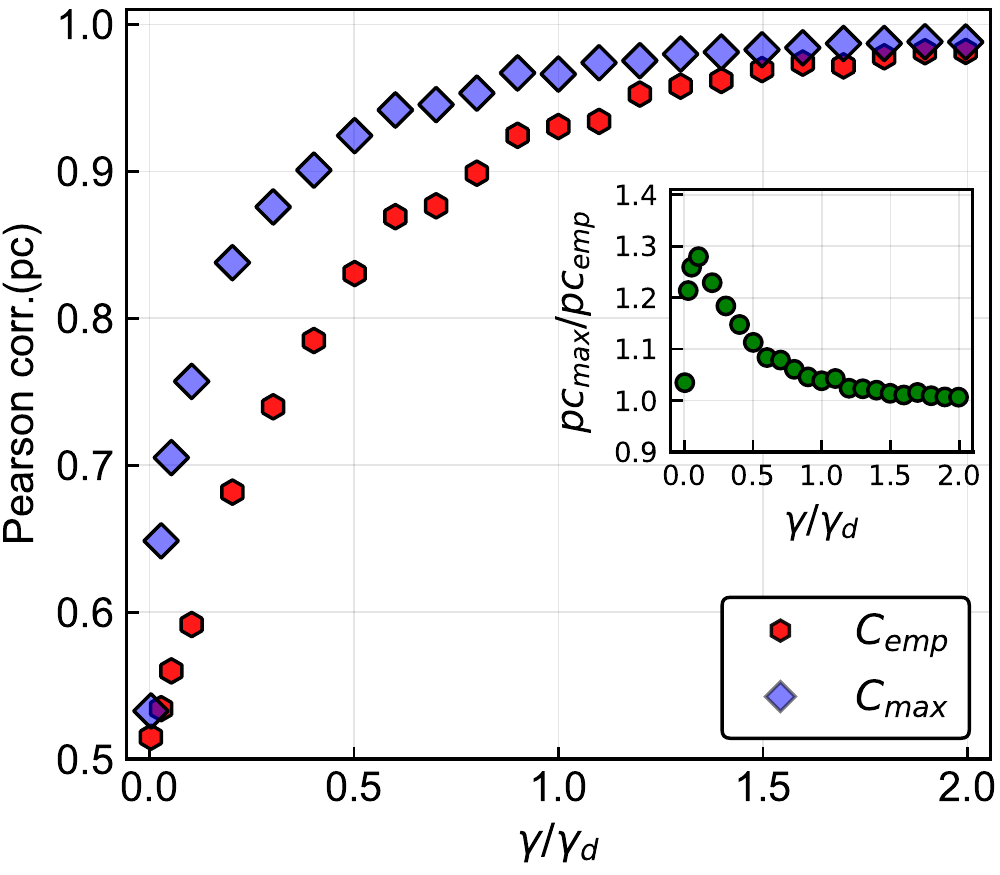}
	\caption{ Pearson correlation between empirical /maximum-likelihood   covariance matrices and the true covariance matrix. The inset plot represents the ratio between the Pearson correlation for the  maximum-likelihood covariance matrix and the one for the empirical covariance matrix. Simulations are performed for a tree of $512$ leaves  and  system size $L=4$.}
	\label{fig:pears_L4}
\end{figure}

\begin{figure*}[!htb]
	%	\begin{subfigure}{}
			\centering\includegraphics[keepaspectratio=true,width=1.0\textwidth]{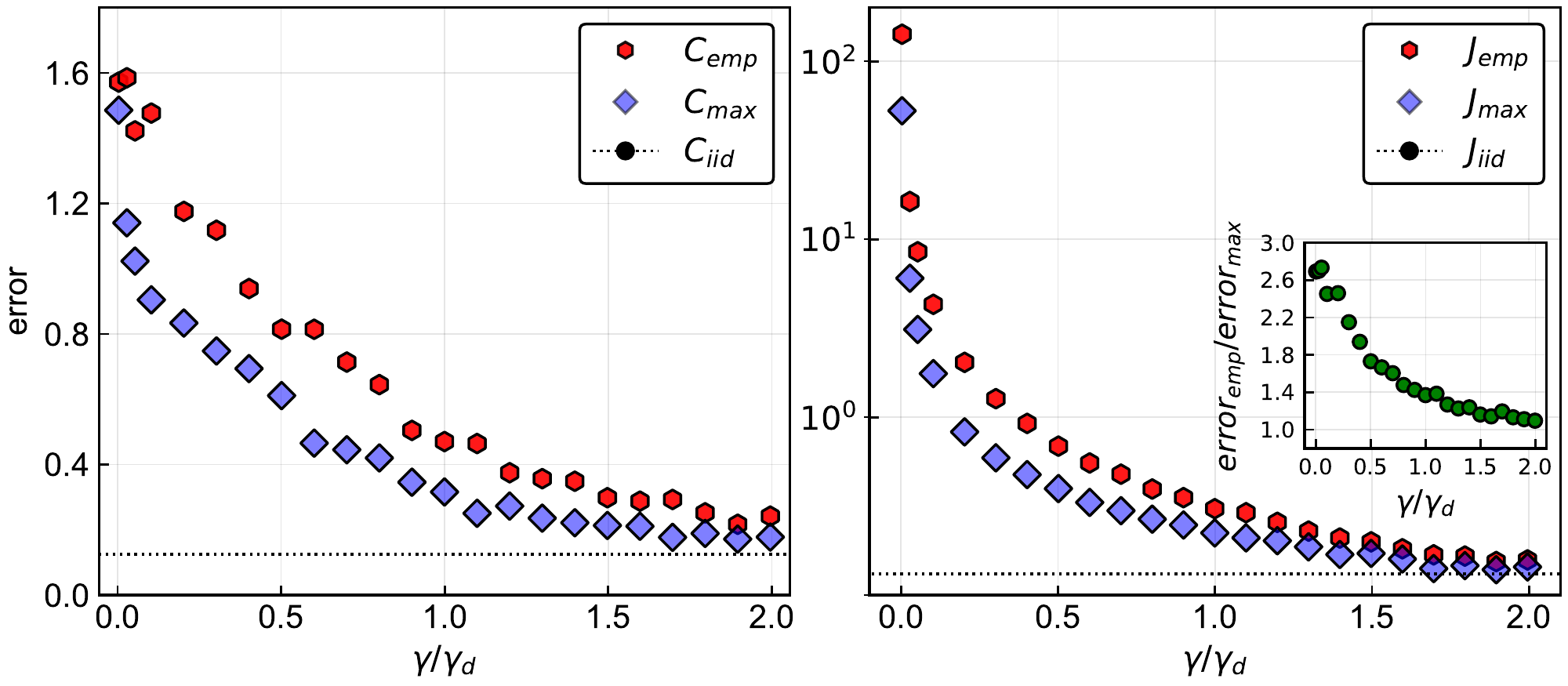}
	%	\end{subfigure}
		\hspace{1mm}
	%	\begin{subfigure}{}
	%		\centering\includegraphics[keepaspectratio=true,width=0.45\textwidth]{Figures/epsilon_error_J_L4_balanced_tree_100.pdf}
	%	\end{subfigure}
	\caption{\textbf{Left:}Relative $l2$-error between empirical or maximum-likelihood  covariance matrices and the true covariance matrix.  \textbf{Right:}Relative $l2$-error between empirical /maximum-likelihood  coupling matrices and the true coupling matrix. Logarithmic scale is chosen for the $y$-axis because of large values of the error at low $\gamma$. The inset in both panels show the ratio between the two errors.}
	\label{fig:error_1_L4}
\end{figure*}

For each repetition of the sampling process, we perform our maximum likelihood procedure and obtain an inferred covariance matrix $\bm{C}_{max}$. 
As a means of comparison, we also compute the empirical covariance matrix $\bm{C}_{emp}$ as if leaf configurations were independent. 
Fig.~\ref{fig:pears_L4} shows the Pearson correlation between the real covariance matrix $\bm{C}=\bm{J}^{-1}$ and the empirical or inferred ones in the $L=4$ case (similar figures for $L=10$ are in Appendix \ref{sub:supplementary_figures}). 
As expected, both methods perform well in the large $\gamma$ limit with a correlation close to $1$, and worse in the low $\gamma$ limit. 
In this latter case, correlations due to phylogeny are too strong for our maximum likelihood method to pick up signal, and both methods perform equally poorly. 
However, there exists an intermediate regime where $\bm{C}_{max}$ is much closer to the actual correlation than  $\bm{C}_{emp}$. 
In Fig.~\ref{fig:error_1_L4}, we plot the relative $l2$-error between either covariance matrices in the left panel or coupling matrices in the right panel. 
In both cases, our maximum-likelihood method results in a consistent improvement over the empirical estimator. 
However, the relative error still reaches high values in the low $\gamma$ regime, which is likely due to $\bm{C}_{max}$ and $\bm{C}_{emp}$ being close to singular in this case. 

An interesting way to illustrate the benefits of reconstructing the covariance matrix using knowledge of the tree is to evaluate the gain in {\em effective sample size}. Intuitively, the use of correlated samples reduces the information contained in the data, as compared to an equally large dataset of \emph{i.i.d.} configurations. It is therefore interesting to compare the accuracy of our inferences with the accuracy obtained on smaller but \emph{i.i.d.} samples. To do so, we report in Fig.~\sref{fig:error_vs_nseqs} the $l2$-error between true and empirical covariances computed from a \emph{i.i.d.} samples of variable sizes $N$. As expected,  the error increases with decreasing values of $N$. 
We can use this in turn to express values of the $l2$-error in correlated samples in terms of effective \emph{i.i.d.} sample sizes.   
For example, the error reached by $\bm{C}_{emp}$ for $\gamma/\gamma_d\in[0.5, 1]$ and $L=4$ corresponds to the one obtained for an \emph{i.i.d.} sample of size $\sim 16$, whereas it corresponds to a sample of size $\sim 32-64$ for $\bm{C}_{max}$.
Thus, our correction is equivalent to increasing by a factor 2-4 the number of effective samples.

Finally, we assess the performance of our method in improving the prediction of the network of interactions between the Gaussian variables $\{x_a\}$. 
We consider that two variables $x_a$ and $x_b$ interact if the corresponding entry in the coupling matrix is non-zero, that is $J_{ab}\neq 0$. 
Using the data, we predict these interactions by taking the largest $n$ elements (in absolute value) of the inferred coupling matrix, resulting in $n$ predictions. 
The fraction TP$/n$ of these $n$ predictions that correspond to non-zero entries in the true matrix (TP = true positives) defines the positive predictive value (PPV). 
This problem is equivalent to the one of predicting contacts in a protein structure 

\begin{figure}[!htb]
	\centering
	\includegraphics[keepaspectratio=true,width=1.0\textwidth]{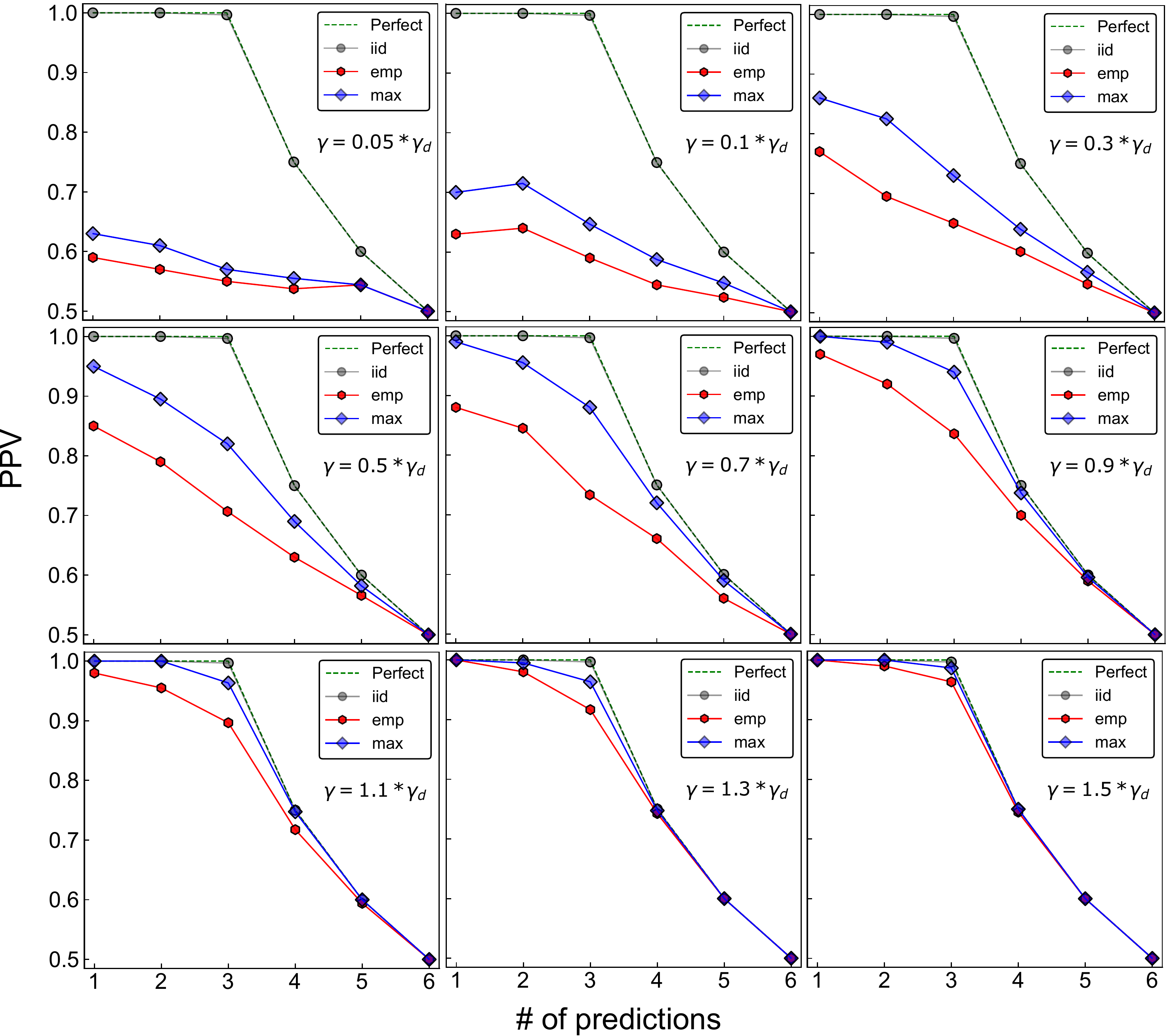}
	\caption{Quality of prediction of interactions for  different values of $\gamma$ and system size $L=4$. Interactions are defined as non-zero elements of the coupling matrix. In the $L=4$ case, there are $6$ possible interactions. Predictions are made by taking the largest elements (in absolute terms) of the inferred coupling matrix. The PPV is the fraction of correctly predicted contacts for a given number of predictions.}
	\label{fig:PPV_L4}
\end{figure}

Fig.~$\ref{fig:PPV_L4}$ shows the PPV as a function of the number of predictions for different values of $\gamma$ and $L=4$ (see Fig. \sref{fig:PPV_L10} for the $L=10$ case). 
In this case, the coupling matrix only has $6$ independent non-diagonal elements, and only 6 predictions can be made. 
Our correction systematically outperforms the predictions from the empirical coupling matrix, with an always larger PPV. 
This gain is negligible in the extreme regimes of very high $\gamma$, where the prediction is close to identical to the one obtained with an \emph{i.i.d.} sample, or very low $\gamma$, where it is essentially random. 
It is however much larger in the intermediate regime, with a significantly improved prediction in the region $\gamma/\gamma_d\in[0.5,1]$.

\section{Discussion}
\label{sec:discussion}

In this work, we proposed a method for inferring parameters of an Ornstein-Uhlenbeck process using data that is correlated through an evolutionary tree. 
We kept a very general setting in which data can in principle represent any set of continuous phenotypic traits or potentially discrete sequences if a continuous approximation is made. 
As such, our approach is purely methodological, and does not directly investigate any particular application. 

We showed that due to the Gaussian and time reversible nature of the OU process, it is possible to write the joint covariance matrix of all data vectors in a simple way. 
The resulting matrix $\mathbb{G}$ consists of block entries that represent covariances between pairs of leaves.
The dependence of these blocks on the coupling matrix $\bm{J}$ characterizing the OU process and on the tree structure can be written explicitly. 
Interestingly, $\mathbb{G}$ only depends on the tree structure through the pairwise path length $\Delta t_{ij}$ separating leaves along the tree.

We then proposed a way to compute the likelihood of the data given the tree and the parameters of the OU process, namely the coupling matrix $\mathbf{J}$ and timescale $\gamma$.
This method relies on computing the eigenvalues and vectors of the joint covariance matrix in an efficient manner. 
Indeed, it is possible to separate this calculation in two steps: the first in which we perform the eigen-decomposition of the matrix $\mathbf{J}$, and the second in which we compute eigenvalues and vectors of matrices $\mathbf{G}^a$ that embed the tree structure. 
This reduces the computational complexity from $\mathcal{O}(L^3N^3)$ for a naive inversion of $\mathbb{G}$ to $\mathcal{O}(L^3) + \mathcal{O}(LN^3)$. 
We also show that this method can be used to compute the gradient of the likelihood with respect to parameters with the same complexity.
This makes the problem of inferring $\mathbf{J}$ amenable to maximum likelihood methods using a gradient ascent approach. 

Finally, we showed that this process gives encouraging results on simulated data, with a more accurate reconstruction of parameters than if empirical estimation was performed. 
These simulations highlight the fact that this method is only useful in the intermediate regime of phylogenetic correlations. 
If the timescale $\gamma$ characterizing the branch lengths of the tree is too large, correlation of data points through the tree is weak and an empirical estimation performs well. 
On the other hand, a very low $\gamma$ results in strong phylogenetic biases that make recovering $\mathbf{J}$ impossible, basically due to a strong reduction of the information in a too redundant dataset. However, in an intermediate regime where intrinsic and historical correlations in the dataset coexist, our tree-aware re-construction of $\mathbf{J}$ results in clear benefits over a tree-unaware empirical estimation. 

A limitation of our approach remains the long computational time. 
Even with the efficient computation of the gradient, it was necessary to use small system sizes, $L=10$ at most, to repeat our inference process many times with simulated data in a reasonable time. 
For this reason, the framework proposed here is limited to a small number of variables. 
In this respect, it is interesting to note that a different manner of computing the likelihood developed in \cite{mitov_fast_2020} and based on Gaussian integrations on every branch of the tree results in an asymptotic complexity of $\mathcal{O}(NL^3)$. 

Although our method can in principle be used for any set of traits, a major motivation in developing it is its potential application to model of proteins sequences. 
Several results in the last years have shown that selection forces shaping the evolution of protein sequences are well described by a pairwise potential. 
The estimation of this potential is performed using homologous sequences, and is therefore biased by the phylogenetic relations between these sequences.  
Results presented here are a first step in disentangling effects due to phylogeny from effects due to selection in a principled way. 

However, there remain several challenges in using this framework for protein sequences. 
First, the computational power required to process actual sequences is much larger than what was needed for the small simulated systems presented here. 
As an example, a protein of length $L=100$ will be represented by $q\times 100=2000$ Gaussian variables, where $q=20$ is the number of amino acids. 
This is of course much larger than the $L=10$ system used as an example to test our approach. 

A second question is the capacity of a continuous variable approximation, necessary when using Ornstein-Uhlenbeck dynamics, to represent dynamical properties of the landscape protein sequences evolve in. 
This type of approximation has been successfully used before, but in quite different contexts \cite{jones_psicov_2012,barton_large_2014,baldassi_fast_2014}. 
Its use in the context of modelling the evolutionary dynamics of protein sequences remains an open question.

\textbf{Acknowledgments:} We acknowledge interesting discussions with Roberto Mulet. PBC and MW acknowlege the hospitality of the Department of  Theoretical  Physics of University  of  Havana, where part of this work was done. Our work was partially funded by the EU H2020 Research and Innovation Programme MSCA-RISE-2016 under Grant
Agreement No. 734439 InferNet.

\clearpage
\bibliography{bib_phylo}

\begin{thebibliography}{10}

\bibitem{Inverse_problem_Berg}
Nguyen~H. Chau, Zecchina~R. N, and Berg J.
\newblock Inverse statistical problems: from the inverse ising problem to data
  science.
\newblock {\em Adv. Phys}, 2017,66,197-261.

\bibitem{levy_potts_2017}
Ronald~M Levy, Allan Haldane, and William~F Flynn.
\newblock Potts {Hamiltonian} models of protein co-variation, free energy
  landscapes, and evolutionary fitness.
\newblock {\em Current Opinion in Structural Biology}, 43:55--62, April 2017.

\bibitem{cocco_inverse_2018}
Simona Cocco, Christoph Feinauer, Matteo Figliuzzi, Remi Monasson, and Martin
  Weigt.
\newblock Inverse {Statistical} {Physics} of {Protein} {Sequences}: {A} {Key}
  {Issues} {Review}.
\newblock {\em Reports on Progress in Physics}, 81(3):032601, March 2018.
\newblock arXiv: 1703.01222.

\bibitem{schneidman2006weak}
Elad Schneidman, Michael~J Berry, Ronen Segev, and William Bialek.
\newblock Weak pairwise correlations imply strongly correlated network states
  in a neural population.
\newblock {\em Nature}, 440(7087):1007--1012, 2006.

\bibitem{roudi2009ising}
Yasser Roudi, Joanna Tyrcha, and John Hertz.
\newblock Ising model for neural data: model quality and approximate methods
  for extracting functional connectivity.
\newblock {\em Physical Review E}, 79(5):051915, 2009.

\bibitem{bialek2012statistical}
William Bialek, Andrea Cavagna, Irene Giardina, Thierry Mora, Edmondo
  Silvestri, Massimiliano Viale, and Aleksandra~M Walczak.
\newblock Statistical mechanics for natural flocks of birds.
\newblock {\em Proceedings of the National Academy of Sciences},
  109(13):4786--4791, 2012.

\bibitem{cavagna2018physics}
Andrea Cavagna, Irene Giardina, and Tom{\'a}s~S Grigera.
\newblock The physics of flocking: Correlation as a compass from experiments to
  theory.
\newblock {\em Physics Reports}, 728:1--62, 2018.

\bibitem{bury2013market}
Thomas Bury.
\newblock Market structure explained by pairwise interactions.
\newblock {\em Physica A: Statistical Mechanics and its Applications},
  392(6):1375--1385, 2013.

\bibitem{borysov2015us}
Stanislav~S Borysov, Yasser Roudi, and Alexander~V Balatsky.
\newblock Us stock market interaction network as learned by the boltzmann
  machine.
\newblock {\em The European Physical Journal B}, 88(12):1--14, 2015.

\bibitem{jaynes1957information}
Edwin~T Jaynes.
\newblock Information theory and statistical mechanics.
\newblock {\em Physical review}, 106(4):620, 1957.

\bibitem{sayers_genbank_2019}
Eric~W Sayers, Mark Cavanaugh, Karen Clark, James Ostell, Kim~D Pruitt, and
  Ilene Karsch-Mizrachi.
\newblock {GenBank}.
\newblock {\em Nucleic Acids Research}, 47(D1):D94--D99, January 2019.

\bibitem{uniprotconsortium_uniprot_2018}
The UniProt Consortium.
\newblock {UniProt}: the universal protein knowledgebase.
\newblock {\em Nucleic Acids Research}, 46(5):2699--2699, March 2018.

\bibitem{durbin1998biological}
Richard Durbin, Sean~R Eddy, Anders Krogh, and Graeme Mitchison.
\newblock {\em Biological sequence analysis: probabilistic models of proteins
  and nucleic acids}.
\newblock Cambridge university press, 1998.

\bibitem{felsenstein_phylogenies_1988}
Joseph Felsenstein.
\newblock Phylogenies and quantitative characters.
\newblock {\em Annual Review of Ecology and Systematics}, 19(1):445--471,
  November 1988.
\newblock Publisher: Annual Reviews.

\bibitem{felsenstein2004inferring}
Joseph Felsenstein and Joseph Felenstein.
\newblock {\em Inferring phylogenies}, volume~2.
\newblock Sinauer associates Sunderland, MA, 2004.

\bibitem{obermayer2014inverse}
Benedikt Obermayer and Erel Levine.
\newblock Inverse ising inference with correlated samples.
\newblock {\em New Journal of Physics}, 16(12):123017, 2014.

\bibitem{rodriguez2019toward}
Edwin Rodriguez~Horta, Pierre Barrat-Charlaix, and Martin Weigt.
\newblock Toward inferring potts models for phylogenetically correlated
  sequence data.
\newblock {\em Entropy}, 21(11):1090, 2019.

\bibitem{qin_power_2018}
Chongli Qin and Lucy~J. Colwell.
\newblock Power law tails in phylogenetic systems.
\newblock {\em Proceedings of the National Academy of Sciences},
  115(4):690--695, January 2018.

\bibitem{horta2020phylogenetic}
Edwin~Rodriguez Horta and Martin Weigt.
\newblock Phylogenetic correlations have limited effect on coevolution-based
  contact prediction in proteins.
\newblock {\em bioRxiv}, 2020.

\bibitem{uhlenbeck_theory_1930}
G.~E. Uhlenbeck and L.~S. Ornstein.
\newblock On the {Theory} of the {Brownian} {Motion}.
\newblock {\em Physical Review}, 36(5):823--841, September 1930.
\newblock Publisher: American Physical Society.

\bibitem{hansen_stabilizing_1997}
Thomas~F. Hansen.
\newblock Stabilizing {Selection} and the {Comparative} {Analysis} of
  {Adaptation}.
\newblock {\em Evolution}, 51(5):1341--1351, 1997.
\newblock \_eprint:
  https://onlinelibrary.wiley.com/doi/pdf/10.1111/j.1558-5646.1997.tb01457.x.

\bibitem{bartoszek_phylogenetic_2012}
Krzysztof Bartoszek, Jason Pienaar, Petter Mostad, Staffan Andersson, and
  Thomas~F. Hansen.
\newblock A phylogenetic comparative method for studying multivariate
  adaptation.
\newblock {\em Journal of Theoretical Biology}, 314:204--215, December 2012.

\bibitem{mitov_fast_2020}
Venelin Mitov, Krzysztof Bartoszek, Georgios Asimomitis, and Tanja Stadler.
\newblock Fast likelihood calculation for multivariate {Gaussian} phylogenetic
  models with shifts.
\newblock {\em Theoretical Population Biology}, 131:66--78, February 2020.

\bibitem{jones_psicov_2012}
David~T. Jones, Daniel W.~A. Buchan, Domenico Cozzetto, and Massimiliano
  Pontil.
\newblock {PSICOV}: precise structural contact prediction using sparse inverse
  covariance estimation on large multiple sequence alignments.
\newblock {\em Bioinformatics}, 28(2):184--190, January 2012.

\bibitem{barton_large_2014}
J.~P. Barton, S.~Cocco, E.~De~Leonardis, and R.~Monasson.
\newblock Large pseudocounts and {L} 2 -norm penalties are necessary for the
  mean-field inference of {Ising} and {Potts} models.
\newblock {\em Physical Review E}, 90(1), July 2014.

\bibitem{baldassi_fast_2014}
Carlo Baldassi, Marco Zamparo, Christoph Feinauer, Andrea Procaccini, Riccardo
  Zecchina, Martin Weigt, and Andrea Pagnani.
\newblock Fast and {Accurate} {Multivariate} {Gaussian} {Modeling} of {Protein}
  {Families}: {Predicting} {Residue} {Contacts} and {Protein}-{Interaction}
  {Partners}.
\newblock {\em PLoS ONE}, 9(3), March 2014.

\bibitem{morcos_direct-coupling_2011}
F.~Morcos, A.~Pagnani, B.~Lunt, A.~Bertolino, D.~S. Marks, C.~Sander,
  R.~Zecchina, J.~N. Onuchic, T.~Hwa, and M.~Weigt.
\newblock Direct-coupling analysis of residue coevolution captures native
  contacts across many protein families.
\newblock {\em Proceedings of the National Academy of Sciences},
  108(49):E1293--E1301, December 2011.

\bibitem{figliuzzi_coevolutionary_2016}
Matteo Figliuzzi, Hervé Jacquier, Alexander Schug, Oliver Tenaillon, and
  Martin Weigt.
\newblock Coevolutionary {Landscape} {Inference} and the {Context}-{Dependence}
  of {Mutations} in {Beta}-{Lactamase} {TEM}-1.
\newblock {\em Molecular Biology and Evolution}, 33(1):268--280, January 2016.

\bibitem{russ_evolution-based_2020}
William~P. Russ, Matteo Figliuzzi, Christian Stocker, Pierre Barrat-Charlaix,
  Michael Socolich, Peter Kast, Donald Hilvert, Remi Monasson, Simona Cocco,
  Martin Weigt, and Rama Ranganathan.
\newblock An evolution-based model for designing chorismate mutase enzymes.
\newblock {\em Science}, 369(6502):440--445, July 2020.
\newblock Publisher: American Association for the Advancement of Science
  Section: Report.

\bibitem{singh2017multiOU}
Rajesh Singh, Dipanjan Ghosh, and R.~Adhikari.
\newblock Fast bayesian inference of the multivariate ornstein-uhlenbeck
  process.
\newblock {\em arxiv:1706.04961}, 2017.

\bibitem{Raffenetti1970GEA}
Richard~C. Raffenetti and Klaus. Ruedenberg.
\newblock Parametrization of an orthogonal matrix in terms of generalized
  eulerian angles.
\newblock {\em International Journal of Quantum Chemistry, Vol.III s,625-634},
  1970.

\bibitem{Shepard_param_OM}
Ron Shepard, Scott~R. Brozell, and Gergely Gidofalvi.
\newblock The representation and parametrization of orthogonal matrices.
\newblock {\em Journal of Physical Chemistry A, 119,7924-7939}, 2015.

\bibitem{Zygote.jl-2018}
Michael Innes.
\newblock Don't unroll adjoint: Differentiating ssa-form programs.
\newblock {\em CoRR}, abs/1810.07951, 2018.

\bibitem{matrix_cook_book}
Kaare Brandt~Petersen and Michael Syskind~Pedersen.
\newblock {\em The Matrix Cookbook}.
\newblock 2015.

\bibitem{NLopt}
Steven~G. Johnson.
\newblock The nlopt nonlinear-optimization package.
\newblock {\em http://github.com/stevengj/nlopt}.

\bibitem{griewank_automatic_1989}
Andreas Griewank.
\newblock On {Automatic} {Differentiation}.
\newblock In {\em In {Mathematical} {Programming}: {Recent} {Developments} and
  {Applications}}, pages 83--108. Kluwer Academic Publishers, 1989.

\end{thebibliography}
\bibliographystyle{unsrt}

\newpage
\appendix

\section{Description of technical details} % (fold)
\label{sec:supplementary_material}

\subsection{Generating artificial data} % (fold)
\label{sub:generating_artificial_data}

We are interested in the case where the dynamics of the $L$-dimensional Ornstein-Uhlenbeck process takes place on a tree. For example, if configurations $\{\vec{x}\}$ represent quantitative traits of some organisms, the tree can represent the genealogy or phylogeny of these organisms. Therefore, to generate our datasets, we have to be able to simulate the OU process on a tree. In practice, given a rooted tree such as the one shown in Fig.~\ref{fig:sample_tree}  of the main text, we want to sample a configuration $\vec{x}$ for every node in such a way that Eq.~\eqref{eq:OUjoint} holds for every pair of nodes, with time $\Delta t$ being the path length connecting the nodes along the tree. 
	
We use a simple methodology to achieve this. First, note that given an arbitrary configuration $\vec{x}_0$ and a time $\Delta t$, we can generate a new configuration $\vec{x}$ distributed according to the propagator Eq.~\eqref{eq:OUpropagator} by exploiting the transformation
\begin{equation}
	\vx = \Lam^{\Delta t}\vx_0 + \Sig^{1/2}\vec{\eta}\ ,
	\label{eq:OUsample}
\end{equation}
where $\Lam$ and $\Sig$ are defined in Eq.~\eqref{eq:def_lambda}, and $\vec{\eta}$ is a vector of uncorrelated variables drawn individually from the normal distribution $\mathcal{N}(0,1)$. Moreover, if $\vec{x_0}$ is distributed according to the equilibrium distribution Eq.~\eqref{eq:eq_distribution}, then $\vx$ and $\vx_0$ are distributed according to the joint distribution Eq.~\eqref{eq:OUjoint} describing two equilibrium configurations at finite time difference. Note that Eq.~\eqref{eq:OUsample} is quite different from the Langevin Eq.~\eqref{eq:langevin}, which describes the instantaneous dynamics of  $\vx$ in the potential given by $\bm{J}$, and which could also be simulated in more complicated situation where no analytical expression for the propagator can be derived. 
	
Given any already sampled internal node in the tree, Eq.~\eqref{eq:OUsample} allows to emit a configuration for each of its child nodes. To sample the whole tree, we first draw the root configuration $\vx_r$ from the equilibrium distribution Eq.~\eqref{eq:eq_distribution}. By recursive applications of Eq.~\eqref{eq:OUsample}, we then simply work our way down the tree until all leaves are sampled. Only the configurations at the leaves form the data set, and the internal configuration remain hidden to our model-learning task.

 % subsection generating_artificial_data (end)

\subsection{Initializing parameters} % (fold)
\label{sub:initializing_parameters}

\subsubsection{Eigenvalues and eigenvectors of $\mathbf{C}^{-1}$}
\label{ssub:eigenvalues_and_eigenvectors}

We initialize the covariance matrix using the empirical one:
\begin{equation*}
	\mathbf{C}^{emp} = \frac{1}{N}\sum_{i=1}^{N}\vx_i \cdot\vx_i^{\,T}\ .
\end{equation*}
Its eigenmodes $\{\rho^0_a, \vsa^{\;0}\}$ determine the starting point of the optimization. A suitable parametrization of $\vsa^{\;0}$ in terms of generalized Eulerian angles or a skew symmetric matrix is described below in Sec.~\ref{sub:parametrizations_of_eigenvectors}.
	
	% From empirical covariance matrix $c^*=\frac{1}{N}\sum_{n=1}^{N}\mathbf{x}_n \mathbf{x}_n^T$ we could extract information  to  initialize the parameters in the optimization process. 

	% The eigenvalues of empirical couplings $j^*=1/c^*$  will define $\rho^0_i$ ands its eigenvectors  an orthogonal matrix $\bm S^0$ which satisfy the equations systems:
	% \begin{equation}
	% S_{ij}(\theta^0)=S^0_{ij}
	% \label{recurrent_eq}
	% \end{equation}
	% to find the orthogonal parameters which reproduce the initial orthogonal matrix, equation \ref{recurrent_eq} must be inverted or equivalently the solution set $\bm \theta^0$ could be numerically obtained minimizing the function:

	% \begin{equation}
	% f(\bm \theta^0)=\sum_{i,j}\left[ S^0_{ij}-S_{ij}(\bm \theta^0)\right] ^2
	% \label{obtain_theta0}
	% \end{equation}

	% subsubsection eigenvalues_and_eigenvectors (end)

\subsubsection{Time scale parameter $\gamma$}

The optimization also requires that we initialize the time scale $\gamma$. For coherence with the last section, we need to find the optimal $\gamma$ given the data $\mathbf{X}$, the tree, and the OU process defined by the empirical covariance matrix. 

The probability distribution $P$ for the configurations of two leaves $\vx_i$ and  $\vx_j$  separated by time $\Delta t_{ij}$ is given by Eq.~\eqref{eq:OUjoint} of the main text. With this distribution we can analytically calculate the average of the scalar product $\vx_i^{\,T}\cdot\vx_j$ of two equilibrium configurations at given time separation:
% \begin{align}
% \langle \vx_i^{\,T}\cdot\vx_j \rangle_P &= \int \ddroit \vx_i \ddroit \vx_j P(\vx_i,\vx_j|\Delta t_{ij}) \sum_{a=1}^L   x_i^{a}x_j^{a}  \nonumber \\
% 	&=\sum_{a=1}^L \langle  x_i^{a} x_j^a \rangle_{P}.
% \end{align}
\begin{equation}
	\langle \vx_i^{\,T}\cdot\vx_j \rangle_P = \sum_{a=1}^L \langle  x_i^{a} x_j^a \rangle_{P}.
\end{equation}
The covariance $\langle  x_i^{a} x_j^a \rangle_{P}$ of two observations separated by time $\Delta t_{ij}$ is given by Eq.~\eqref{eq:pairwisecov}. Using this, we find
\begin{align}
		\nonumber\langle \vx_i^{\,T}\cdot\vx_j \rangle_P &= \sum_{a=1}^L \left( \Lam^{\Delta t_{ij}} \mathbf{C} \right)_{aa}\\
		\nonumber&= \Tr\left( \Lam^{\Delta t_{ij}} \mathbf{C} \right)\\
		&= \sum_{a=1}^L \rho_a^{-1}e^{-\gamma\rho_a\Delta t_{ij}}.
		\label{eq:scalar_product}
\end{align}
Having initialized the covariance matrix $\mathbf{C}$ with its empirical value, we know the values of all members of the r.h.s. of Eq.~\eqref{eq:scalar_product} except the one of $\gamma$. To find an initial value of $\gamma$ which is consistent with the data and the empirical covariance matrix for all pairs of data configurations $i<j$, we search for one that best explains the observed scalar products between configurations. We thus define $\gamma^0$ to be the argument minimizing the functional $F(\gamma)$: 
\begin{equation}
    F(\gamma) = \sum_{1\leq i<j\leq N}\left[ \vx_i^{\,T}\cdot\vx_j - \sum_{a=1}^L \rho_a^{-1}e^{-\gamma\rho_a\Delta t_{ij}} \right].
\end{equation} 
Since $F$ depends on a single scalar parameter, it is straightforward to minimize it and thereby to initialize $\gamma$ to an empirically reasonable value.

\subsection{Parametrizations of  eigenvectors } % (fold)
\label{sub:parametrizations_of_eigenvectors}
\subsubsection{Parametrization using generalized Eulerian angles}
The idea is to write the base vectors $\vec{s}_a$ as columns of an orthogonal matrix $\bm{T}$, and to parameterize this matrix in terms of $L(L-1)/2$ independent variables  $\theta_ {pq}$ with $1 \leq p <q \leq L $. These parameters are called generalized Eulerian angles, since they generalize Eulerian angles to $L>3$. 

To construct this matrix we start from a rotational transformation in a two-dimensional subspace of an $L$-dimensional space. It is given by an $L$-dimensional matrix of the form
\begin{equation} 
\bm a_{pq} =  \left(
\begin{array}{cccccc}
1 &   &  &  &  &\\
 & 1& & & & \\
& & \cos\theta_{pq} & & \sin\theta_{pq}\\
& &  & 1 & &\\
& &  -\sin\theta_{pq}& & \cos\theta_{pq} &\\
& &  & &  & 1\\
\end{array}
\right) \ ,
\label{EP_eq:1}
\end{equation}
where all diagonal elements are unity except for the diagonal elements in the $p$th and the $q$th column, which equal $\cos\theta_{pq}$. All off-diagonal elements are zero except for the one corresponding to the intersection of the $p$th row and the $q$th column, which is $\sin\theta_{pq}$, and that on the intersection of the $q$th row and the $p$th column, which equals $-\sin\theta_{pq}$. There are $L(L - 1)/2$  matrices of this form, corresponding to all choices of $p$ and $q$ with $1\leq p <q\leq L$. 

An arbitrary $L$-dimensional orthogonal matrix $\bm T$ can be represented as a product of these $L(L-1)/2$ orthogonal matrices with appropriate values of the $L(L-1)/2$ independent parameters $\theta_{pq}$. Ref. \cite{Raffenetti1970GEA} exposes a recursive algorithm to efficiently perform the matrix multiplication, as well as the construction of the derivatives of $\bm T$ with respect to parameters $\theta_{pq}$. The main equations are presented below.

The matrix multiplication  can be done by a sequence of $L$ steps  implied by the following recurrence relations where $n$ goes from 1 to $L$:
\begin{equation}
	\bm T=\bm T^{(L)},
	\label{EP_eq:2}
\end{equation}

\begin{equation}
\bm T^{(n)}= \bm A^{(n)} \bm t^{(n)},
\label{EP_eq:3}
\end{equation}

\begin{equation} 
\bm t^{(n)} =  \left(
\begin{array}{cc}
\bm T^{(n-1)} & \bm 0  \\
\bm 0 &  \bm I\\
\end{array}
\right),
\label{EP_eq:4}
\end{equation}

\begin{equation} 
\bm T^{(1)}=1,
\label{EP_eq:init}
\end{equation}

\begin{equation}
A^{(n)}= \bm a_{n,n}\,\bm a_{n-1,n}\dotsc\bm a_{2,n} \,\bm a_{1,n} ,
\label{EP_eq:5}
\end{equation}

where the $\bm a_{n,m}$ matrices are defined by \eqref{EP_eq:1} for $n\neq m$, and $\bm a_{n,n}$ is the identity matrix of dimension $n$.

The recurrence equations given by \eqref{EP_eq:3},  can be explicitly written as 

\begin{equation}
T^{(n)}_{kl}=\cos\theta_{kn} \cdot t^{(n)}_{kl}-\sin\theta_{kn}\cdot z^{(n)}_{kl}\;\;\; \text{with} \;\;\; k,l=1,...,n
\label{EP_eq:6}
\end{equation}

where

\begin{equation}
z^{(n)}_{kl} =
\begin{cases}
\delta_{ln} & \text{for $k=1$}\\
\sin\theta_{k-1 n}\cdot t^{(n)}_{k-1 l}+\cos\theta_{k-1 n}\cdot z^{(n)}_{k-1 l} & \text{for $k=2,...,n$}  
\end{cases}
\label{EP_eq:7}
\end{equation}
with $\theta_{nn}=\pi/2$.

Thus, if $\bm T^{(n-1)}$ is given, we find $\bm t^{(n)}$ from equation \eqref{EP_eq:4}. Then from elements $t_{kl}^{(n)}$ we get $z_{kl}^{(n)}$ using \eqref{EP_eq:7} and finally from $z_{kl}^{(n)}$ and $t_{kl}^{(n)}$ we obtain $T^{(n)}_{kl}$. 

Therefore eigenvectors $\vec{s}_a$  can be chosen as $a$th column  of matrix $\bm{T}$:

\begin{equation}
s^{k}_a=T^{(L)}_{.,a}=\cos\theta_{kL} \cdot t^{(L)}_{ka}-\sin\theta_{kL}\cdot z^{(L)}_{ka} \;\;\; \text{for} \;k=1,..,L
 \label{EP_eq:8}
\end{equation}

\textbf{Determination of parameters for a given matrix}

To use this expression, we still need to determine parameters $\bm \theta$  given an orthogonal matrix $\bm T$, such that all equations
\begin{equation}
    T_{ij}(\bm \theta)=T_{ij}
\end{equation}
are satisfied. This system of nonlinear transcendental equations  cannot be solved algebraically. However, it is possible to overcome this issue finding the set of $\bm \theta$ which minimize the square distance between the target and the parametrized matrices: 
\begin{equation}
 {\bm{\hat \theta}} = \text{argmin}_{\bm \theta} \sum_{i<j}\left[ T_{ij}-T_{ij}(\bm \theta)\right] ^2 \ .
\end{equation}
This is useful when we initialize parameters $\bm \theta$ for the matrix formed by the eigenvectors of the empirical covariance matrix.

\textbf{Derivatives with respect to the angular parameters}
 
To compute the derivatives of $\bm T$ with respect to the angular parameters $\theta_{pq}$ we first note that it is possible to rewrite the recurrence step of equation \eqref{EP_eq:3} as the following matrix product
 
 \begin{equation}
 \bm T^{(L)}= \bm B^{(L)} \bm B^{(L-1)}\cdots\bm B^{(3)}\bm B^{(2)}   
 \label{EP_eq:10}
 \end{equation}

where

\begin{equation} 
\bm B^{(n)} =  \left(
\begin{array}{cc}
\bm A^{(n)} & \bm 0  \\
\bm 0 &  \bm I ^{(L-n)}\\
\end{array}
\right)
\label{EP_eq:11}
\end{equation}

is block diagonal, $\bm A^{(n)} $ was defined by Eq.~\eqref{EP_eq:5} and $\bm I ^{(L-n)}$ is unit matrix of $(L-n)$ dimensions. 

From the definition of $\bm A^{(n)} $ we note that the terms $\theta_{pq}$ for $p=1,2,...,q-1$ only occur in the factor $\bm B^{(q)}$ of equation \eqref{EP_eq:10}. 
This allows to write  the derivative of $\bm T$ with respect to $\theta_{pq}$  as the matrix product:

\begin{equation}
 \frac{\partial \bm T}{\partial \theta_{pq}}= \bm B^{(L)} \bm B^{(L-1)}\cdots \frac{\partial \bm B^{(q)}}{\partial \theta_{pq} }\cdots\bm B^{(3)}\bm B^{(2)}   
\label{EP_eq:12}
\end{equation}

where 

\begin{equation} 
 \frac{\partial \bm B^{(q)}}{\partial \theta_{pq} } =  \left(
\begin{array}{cc}
 \frac{\partial\bm A^{(q)}}{\partial \theta_{pq}  } & \bm 0  \\
 \bm 0 &  \bm 0\\
\end{array}
\right)
\label{EP_eq:13}
\end{equation}

Therefore  the calculation of the derivative of $\bm T$ could be done with the following three steps:
\begin{enumerate}
\item Calculate the product $\bm B^{(q-1)}\bm B^{(q-2)}\cdots\bm B^{(3)}\bm B^{(2)}   $
\item  Calculate \begin{equation}
\frac{\partial \bm B^{(q)}}{\partial \theta_{pq} }\bm B^{(q-1)}\cdots\bm B^{(3)}\bm B^{(2)}   = \left(
\begin{array}{cc}
\frac{\partial\bm T^{(q)}}{\partial \theta_{pq}  } & \bm 0  \\
\bm 0 &  \bm 0\\
\end{array}
\right) 
\label{EP_eq:14}
\end{equation}

\item Calculate \begin{equation}
\frac{\partial\bm T}{\partial \theta_{pq}  }=\bm B^{(L)}\bm B^{(L-1)}\cdots \bm B^{(q+1)} \left(
\begin{array}{cc}
\frac{\partial\bm T^{(q)}}{\partial \theta_{pq}  } & \bm 0  \\
\bm 0 &  \bm 0\\
\end{array}
\right) 
\label{EP_eq:15}
\end{equation}

\end{enumerate}

The $q-3$ recurrence steps for step 1  can be carried out using the same recurrence scheme described before for matrix $\bm T$ construction. For step 2, we need to evaluate 
\begin{equation}
\frac{\partial\bm T^{(q)}}{\partial \theta_{pq} }  =
\begin{cases}
- \sin\theta_{pq} \sigma^{(q)}_{kl} & k>p\\
- \sin\theta_{pq} t^{(q)}_{kl} - \cos\theta_{pq} z^{(q)}_{kl} & k=p \\
0 & k<p
\end{cases}
\label{EP_eq:17}
\end{equation}

where the quantities
 \begin{equation}
\sigma^{(q)}_{kl}=\frac{\partial\bm z^{(q)}_{kl}}{\partial \theta_{pq} }  
\label{EP_eq:18}
\end{equation}

can be obtained from \eqref{EP_eq:7}. 
Finally for step 3 we follow $L-q$ recurrence steps described by equations \eqref{EP_eq:6} and \eqref{EP_eq:7} with the two exceptions:

 \begin{equation}
z^{(n)}_{1l}=0\;\;\;\;\;1\leq l < n
\label{EP_eq:19}
\end{equation}
 \begin{equation}
z^{(n)}_{k+1,n}=0  \;\;\;\;\;1\leq k \leq n-1
\label{EP_eq:20}
\end{equation}

\subsubsection{Parametrization in term of the exponential of a skew-symmetric matrix }
The exponential of a skew-symmetric matrix   $\bm X=-\bm X^T$ is  a  special orthogonal  matrix  : 
%Any orthogonal matrix can be written as the exponential of a skew-symmetric matrix $\bm X=-\bm X^T$: 
\begin{equation}
	\label{eq:sksym_def}
    \bm S=\exp (\bm X).
\end{equation}
This is simply shown by the fact that $\exp(\bm X)^T=\exp(\bm X^T)=\exp(-\bm X)=\exp(\bm X)^{-1}$ and $\det (\bm S)=\exp{(\Tr \bm X)}=1$ since $\Tr \bm X=0$ for a skew-symmetric matrix. Furthermore, it is always possible to obtain a skew-symmetric matrix $\bm X$ from a special orthogonal matrix $\bm S$ by inverting the exponential relation, $\bm X=\log \bm S$ \cite{Shepard_param_OM}. 

The advantage of expressing $\bm S$ in this form is that $\bm X$ has $L(L-1)/2$ entries that can be varied independently. 
This allows us to perform the optimization over $L(L-1)/2$ independent parameters, with derivatives with respect to the independent entries of $\bm X$ being defined by
\begin{equation}
	\frac{\partial \bm S}{\partial X_{jk}}=\lim_{h\rightarrow 0} \frac{1}{h} \left( \exp(\bm X+ h \bm E^{jk}) - \exp(\bm X)\right) 
	\label{eq:derivative_sksym}
\end{equation}
where $\bm E^{jk}$ for $j > k$ is defined as a skew-symmetric matrix that has only two nonzero entries in positions $(j,k)$ and $(k,j)$:
\begin{equation}
    E^{pq}_{jk}=\delta_{pj}\delta_{qk}-\delta_{pk}\delta_{qj}
\end{equation}

It is not possible to give a simple analytical form to Eq.~\eqref{eq:derivative_sksym}. 
However, since $\bm S$ is obtained through a simple algebraic expression (Eq.~\eqref{eq:sksym_def}), it is possible to compute its derivative with respect to entries of $\bm X$ through automatic differentiation techniques \cite{griewank_automatic_1989}. 
We implemented this process using the Julia package Zygote.jl \cite{Zygote.jl-2018}.

 \subsection{Homogeneous and fully balanced tree}
 \label{sub:homogeneous_and_fully_balanced_tree}

Let's assume that the tree is binary, symmetric and completely homogeneous with all branches having the same length $\Delta t$. 
As an example, the covariance matrix for such a tree with $K=2$ levels with branching and four leaves is
\begin{equation} 
\mathbb{G} =  \left(
\begin{array}{cccc}
{ \bm C} & {\bm C} \Lambda ^{2\Delta t} & {\bm C} \Lambda ^{4\Delta t} & {\bm C} \Lambda ^{4\Delta t} \\
{\bm C} \Lambda ^{2\Delta t} & {\bm C} & {\bm C} \Lambda ^{4\Delta t} & {\bm C} \Lambda ^{4\Delta t} \\
{\bm C} \Lambda ^{4\Delta t} & {\bm C} \Lambda ^{4\Delta t} & {\bm C} & {\bm C} \Lambda ^{2\Delta t} \\
{\bm C} \Lambda ^{4\Delta t} & {\bm C} \Lambda ^{4\Delta t} & {\bm C} \Lambda ^{2\Delta t} & {\bm C} \\
\end{array}\label{eq:G}
\right).
\end{equation}

The associated matrix $\bm G^a=z(\rho_a,\gamma,\Delta t)$  defined in Eq.~\eqref{eq:G} becomes

% \begin{equation} 
% \bm G^a =  \left(
% \begin{array}{cccc}
% {  \rho^{-1}_a } & {\rho^{-1}_a } e ^{-2\gamma\rho_a\Delta t} & {\rho^{-1}_a } e ^{-4\gamma\rho_a\Delta t}& {e ^{-4\gamma\rho_a\Delta t}} \\
% {\rho^{-1}_a } e ^{-2\gamma\rho_a\Delta t} & {\rho^{-1}_a } & {\rho^{-1}_a } e ^{-4\gamma\rho_a\Delta t} & {\rho^{-1}_a } e ^{-4\gamma\rho_a\Delta t} \\
% {\rho^{-1}_a } e ^{-4\gamma\rho_a\Delta t} & {\rho^{-1}_a} e ^{-4\gamma\rho_a\Delta t}& {\rho^{-1}_a } & {\rho^{-1}_a } e ^{-2\gamma\rho_a\Delta t}\\
% {\rho^{-1}_a } e ^{-4\gamma\rho_a\Delta t} & {\rho^{-1}_a } e ^{-4\gamma\rho_a\Delta t} & {\rho^{-1}_a } e ^{-2\gamma\rho_a\Delta t} & {\rho^{-1}_a } \\
% \end{array}\label{eq:G_a}
% \right).
% \end{equation}
\begin{equation} 
\bm G^a =  \rho_a^{-1}\left(
\begin{array}{cccc}
1 & e ^{-2\gamma\rho_a\Delta t} & e ^{-4\gamma\rho_a\Delta t}& {e ^{-4\gamma\rho_a\Delta t}} \\
e ^{-2\gamma\rho_a\Delta t} & 1 & e ^{-4\gamma\rho_a\Delta t} & e ^{-4\gamma\rho_a\Delta t} \\
e ^{-4\gamma\rho_a\Delta t} & e ^{-4\gamma\rho_a\Delta t}& 1 & e ^{-2\gamma\rho_a\Delta t}\\
e ^{-4\gamma\rho_a\Delta t} & e ^{-4\gamma\rho_a\Delta t} & e ^{-2\gamma\rho_a\Delta t} & 1 \\
\end{array}\label{eq:G_a}
\right).
\end{equation}

Matrices such as the one in \eqref{eq:G_a} are called \emph{hyper-geometric}.
For dimensions $2^K$, they have $K+1$ different eigenvalues given by: 

\begin{equation}
\label{eq:lambda_hyper}
\lambda_k(\rho_a,\gamma) =\rho^{-1}_a *
\begin{cases}
 1+\sum_{l=1}^{k-1} 2^{l-1} e ^{-2l\gamma\rho_a\Delta t}-2^{k-1} e ^{-2k\gamma\rho_a\Delta t}, \;\text{for}\; k\in[1,K] \\
 1+\sum_{l=1}^{K} 2^{l-1} e ^{-2l\gamma\rho_a\Delta t}, \;\text{for}\; k=K+1
\end{cases}
\end{equation}

where $\lambda_{K+1}\ge\lambda_{K}\cdots\ge\lambda_{1}$. 
For $k<K+1$, the degeneracy of eigenvalue $\lambda_k$ is $d_k=2^{K-k}$.  
The associated eigenvectors are independent of the parameter $\rho_a$ and reflect the events in the phylogenetic tree.
Each eigenvector $\vec{u}_k$ of length $2^K$ captures the duplication events in the $(K+1 -k) st$ generation:

\begin{equation} 
\nonumber
 \vec{u}_k =
\begin{cases}
{(\underbrace{\overbrace{1,\ldots,1}^{2^{k-1}},\overbrace{-1,\ldots,-1}^{2^{k-1}}}_Q,0,\ldots,0)\ \bigcup \Gamma(u_k) }, \;\text{for}\; k\in[1,K] \\
(1,1,1,\ldots,1,1,1), \;\text{for}\; k=K+1
\label{eq:eigenvec_simpler_case}
\end{cases}	
\end{equation}

where $\Gamma(\vec{u}_k )$ represents the $d_k$ combinations obtained by shifting the block of length $Q=2^k$, generating all eigenvectors corresponding to the eigenvalue $\lambda_k$.  
The eigenvectors are orthogonal to each other, and can be  normalized and arranged horizontally into a matrix $U$. 

To compute the gradient of the likelihood, derivatives of $\lambda_k(\rho_a,\gamma)$ with respect to $\rho_a$ and $\gamma$ can be directly obtained from expression \eqref{eq:lambda_hyper}.

\subsection{Optimization scheme}
\label{sub:optimization_scheme}

The proposed inference scheme was transformed into a multidimensional nonlinear optimization problem for which we can compute the gradient of the optimized quantity $\mathcal{L}$.
To solve it, we used a variant of the quasi-Newton methods (QNM).  
The main feature in QNM when compared to standard Newton method is that the Hessian matrix $\bm H$ is approximated instead of computed exactly. 
When maximizing the likelihood $\mathcal{L}$ with respect to parameters $\vec{\theta}$, the direction of the change of parameters $\Delta\vec{\theta}$ is determined by

$$ \Delta\vec{\theta} = \mathbf{\hat{H}}_k \nabla \mathcal{L}(\vec{\theta}_k), $$

where $\vec{\theta}_k$ and $\mathbf{\hat{H}}_k$ respectively represent the parameter values and the approximation of the Hessian at the $k$-th iteration. 
% $$\nabla \mathcal{L}(\vec{\theta}_k+\Delta\vec{\theta})=\nabla \mathcal{L}(\vec{\theta}_k)+\mathbf{\hat{H}} \Delta\vec{\theta} $$,
Various QNMs differ in their approximation of the Hessian matrix.
We used the LBFGS variant which chooses $\mathbf{\hat{H}}_k$ as a positive definite matrix where

$$\mathbf{\hat{H}}_{k+1}=\mathbf{\hat{H}}_{k}+\frac{y_k y^{T}_k}{y_k\Delta \theta_k} - \frac{\mathbf{\hat{H}}_k\Delta \theta_k(\mathbf{\hat{H}}_k\Delta \theta_k)^T}{\Delta \theta^T_k\mathbf{\hat{H}}_k\Delta \theta_k} \;\;\;\text{with}\;\;\; y_k=\nabla \mathcal{L}(\vec{x}_k+1)-\nabla \mathcal{L}(\vec{x}_k).$$

Implementation of the method was done using NLopt package \cite{NLopt} .
 
\subsection{Supporting figures}
\label{sub:supplementary_figures}

\begin{figure}[!htb]
	\centering
	\includegraphics[keepaspectratio=true,width=0.5\textwidth]{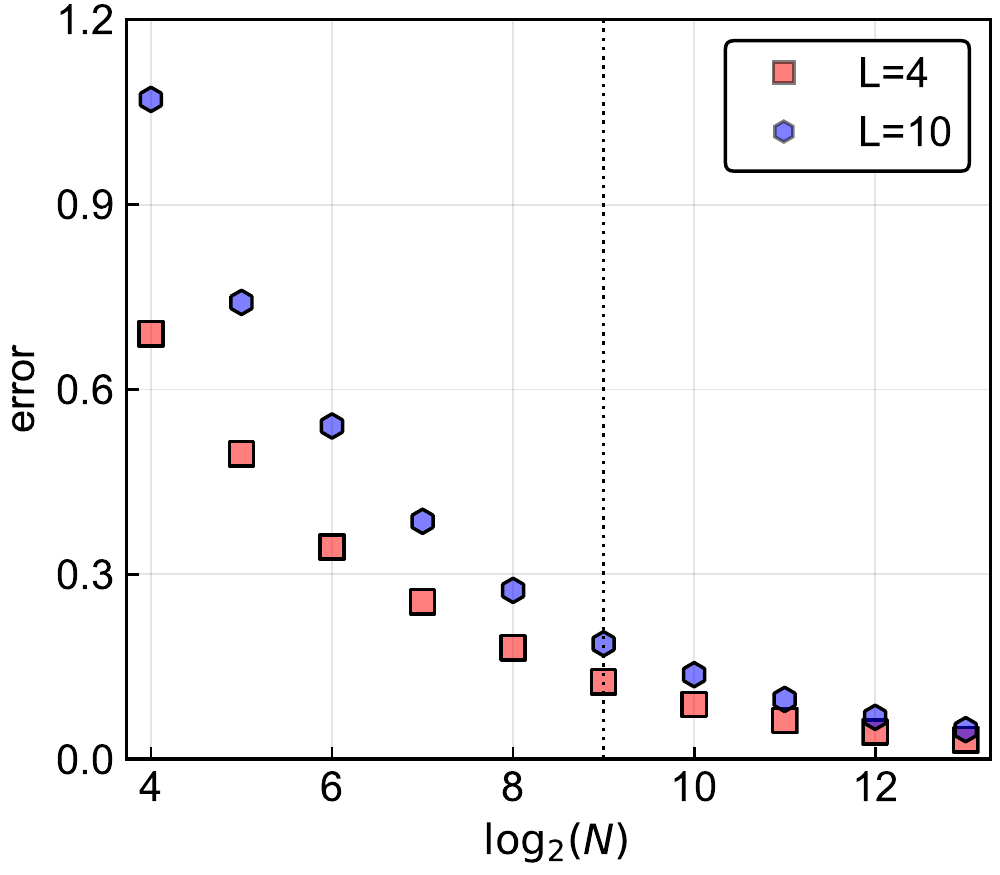}
	\caption{Relative $l2$-error between the empirical covariance matrix calculated from an \emph{i.i.d.} sample and the true covariance matrix, for system sizes $L=4$ and $L=10$. The dashed vertical line corresponds to the number of leaves of the tree used in the simulations.}
	\label{fig:error_vs_nseqs}
\end{figure}

\begin{figure}[!htb]
	\centering
	\includegraphics[keepaspectratio=true,width=0.5\textwidth]{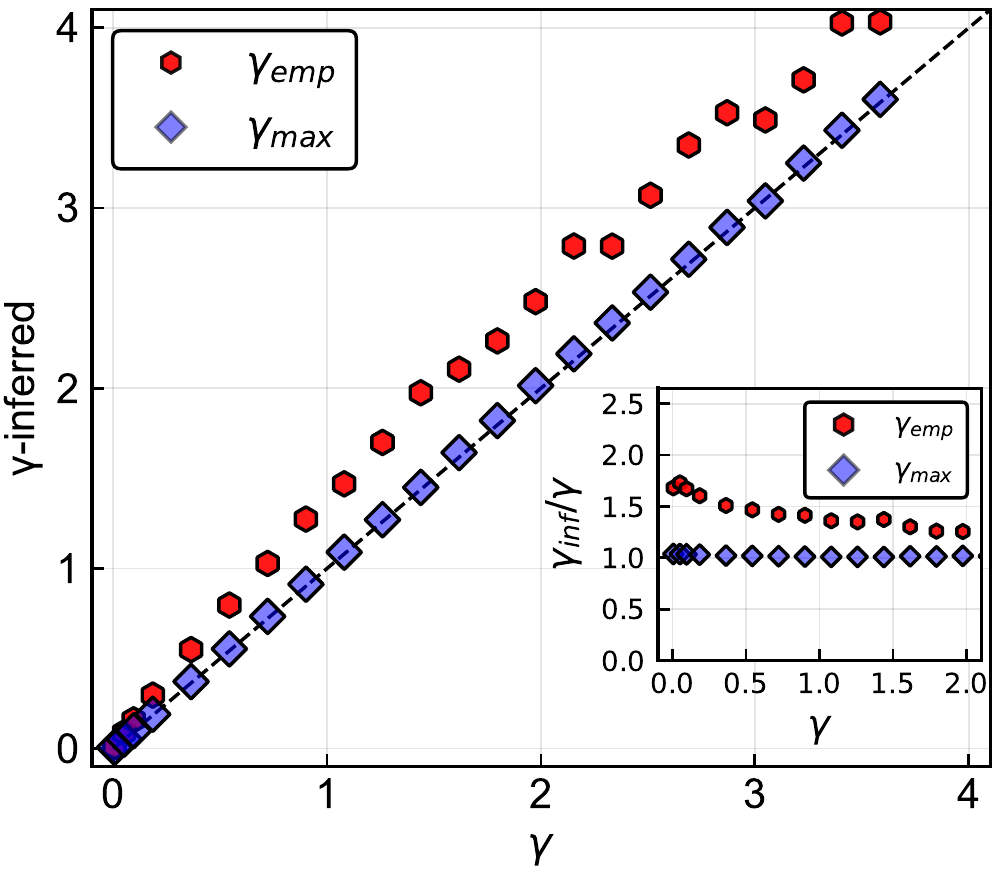}
	\caption{Inferred $\gamma$ values as a function of real $\gamma$, for system size $L=4$. $\gamma_{emp}$ is the value obtained by the process described in section \ref{sub:initializing_parameters} of the appendix. $\gamma_{max}$ is the value inferred by the maximum-likelihood calculation. 
	The inset represents the ratio of both inferred parameters $\gamma_{emp}$ or $\gamma_{max}$ to the real $\gamma$.}
	\label{fig:scatter_gamma_L4}
\end{figure}

%\begin{figure}[!htb]
%	\centering
%	\includegraphics[keepaspectratio=true,width=0.5\textwidth]{Figures/Pearson_correlation_rate_full_range.pdf}
%	\caption{  Ratio between Pearson correlation of the  maximum-likelihood  and true covariance  matrix to the Pearson correlation of the  empirical and true covariances matrices for the two system sizes $L=4$ and $L=10$.}
%	\label{fig:pearson_corr_comp_L4}
%\end{figure}

\begin{figure}[!htb]
	\centering
	\includegraphics[keepaspectratio=true,width=0.5\textwidth]{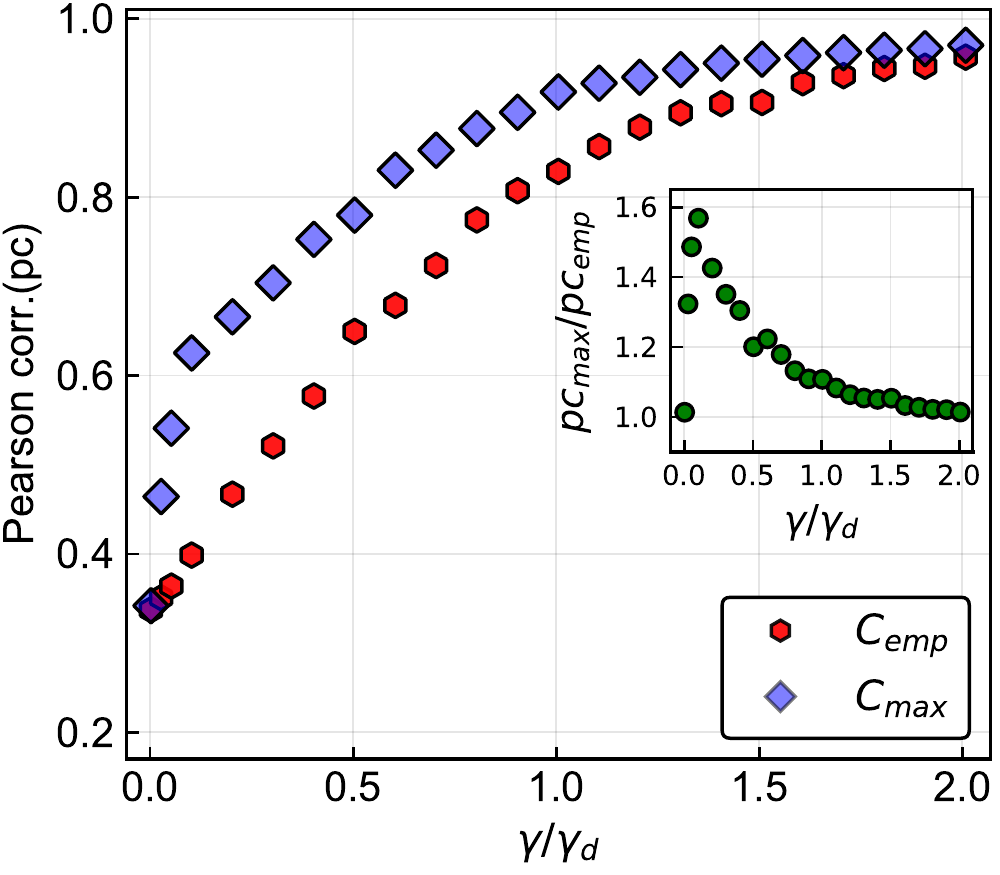}
	\caption{ Pearson correlation between empirical /maximum-likelihood   covariance matrices and the true covariance matrix, the inset plot represent the ratio between the person correlation for the  maximum-likelihood covariance matrix and the one for the empirical covariance matrix.}
	\label{fig:pears_L10}
\end{figure}

\begin{figure}[!htb]
	\centering
	\includegraphics[keepaspectratio=true,width=0.5\textwidth]{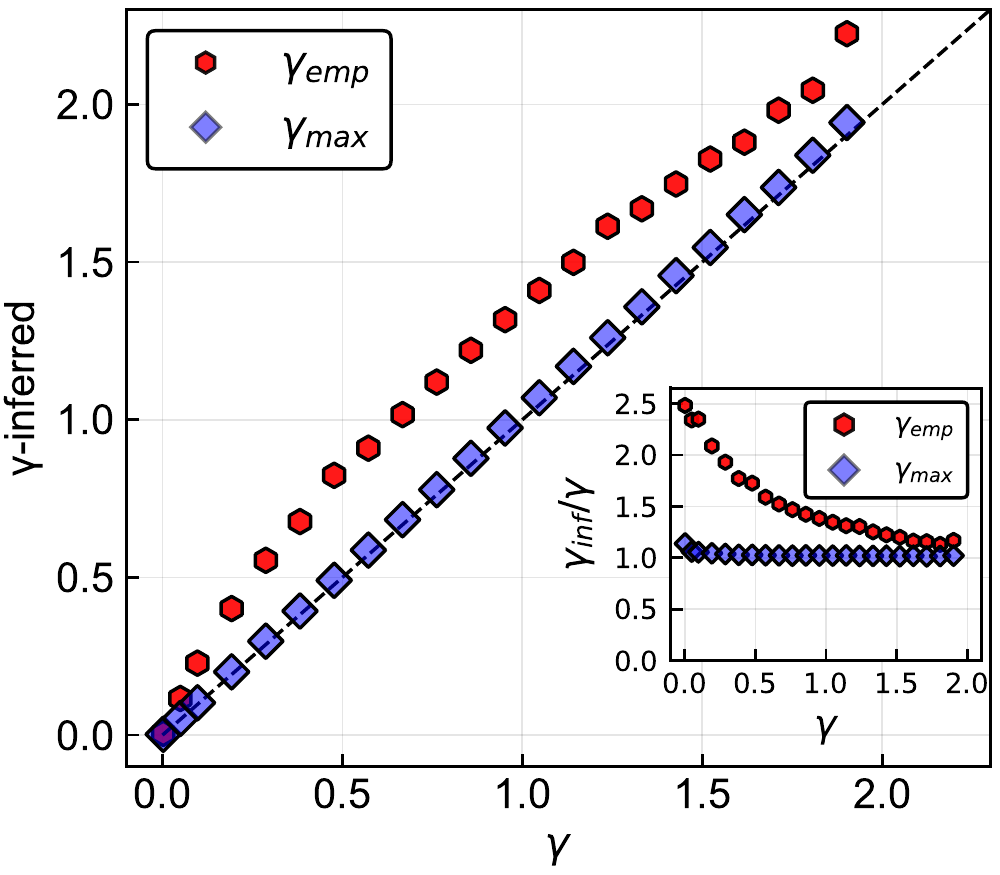}
	\caption{Inferred $\gamma$ values as a function of real $\gamma$, for system size $L=10$. $\gamma_{emp}$ is the value obtained by the process described in section \ref{sub:initializing_parameters} of the appendix. $\gamma_{max}$ is the value inferred by the maximum-likelihood calculation. 
	The inset represents the ratio of both inferred parameters $\gamma_{emp}$ or $\gamma_{max}$ to the real $\gamma$. }
	\label{fig:scatter_gamma_L10}
\end{figure}

 \begin{figure*}[!htb]
 	%	\begin{subfigure}{}
 			\centering\includegraphics[keepaspectratio=true,width=1.0\textwidth]{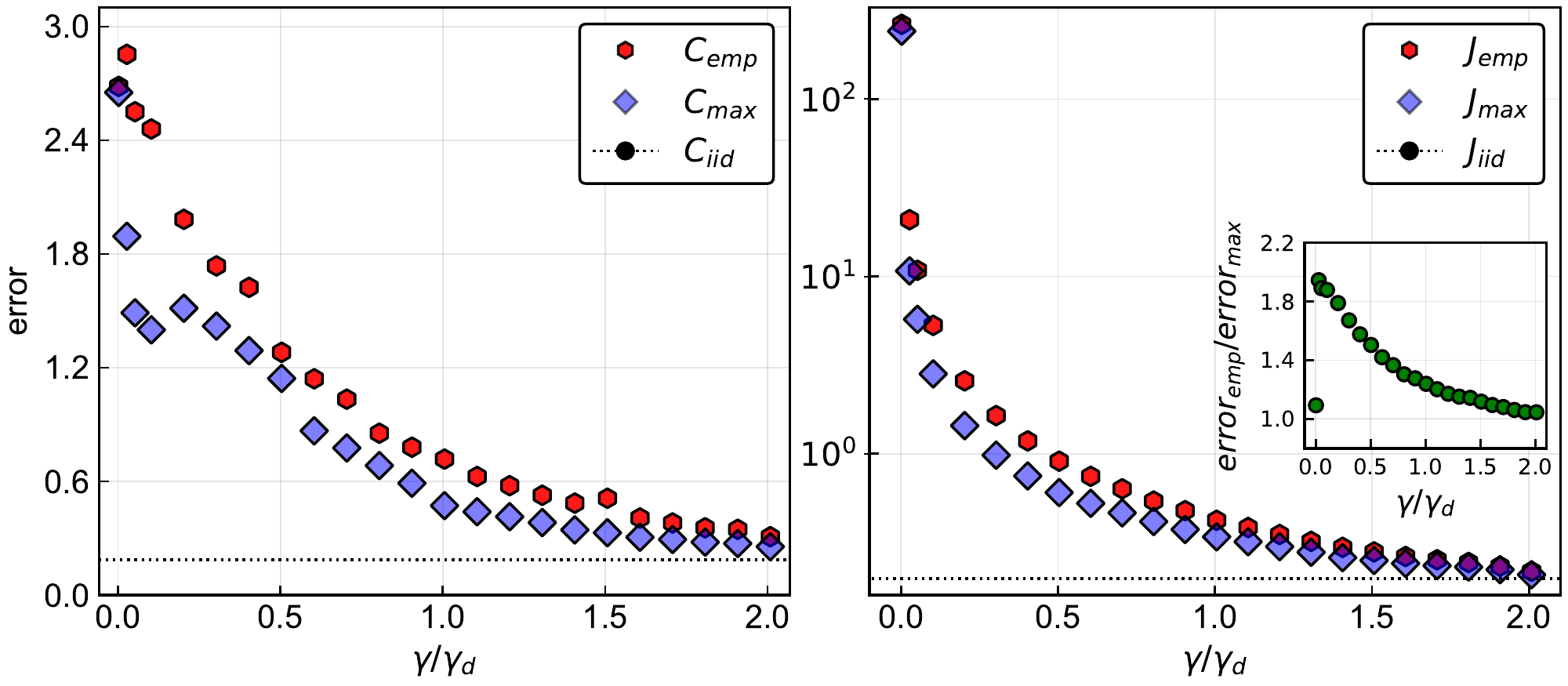}
 %		\end{subfigure}
 		\hspace{1mm}
 %		\begin{subfigure}{}
 %			\centering\includegraphics[keepaspectratio=true,width=0.45\textwidth]{Figures/epsilon_error_J_L10_balanced_tree_100.pdf}
 %		\end{subfigure}
 	\caption{\textbf{Left:} Relative $l2$-error between empirical or maximum-likelihood covariance matrices and the true covariance matrix. \textbf{Right}: Relative $l2$-error between empirical /maximum-likelihood coupling matrices and the true coupling matrix. Logarithmic scale is chosen for the y-axis because of large values of the error at low $\gamma$. The inset in both panels show the ratio between the two errors. For system size $L=10$.}
 	\label{fig:error_1_L10}
 \end{figure*}

\begin{figure}[!htb]
	\centering
	\includegraphics[keepaspectratio=true,width=1.0\textwidth]{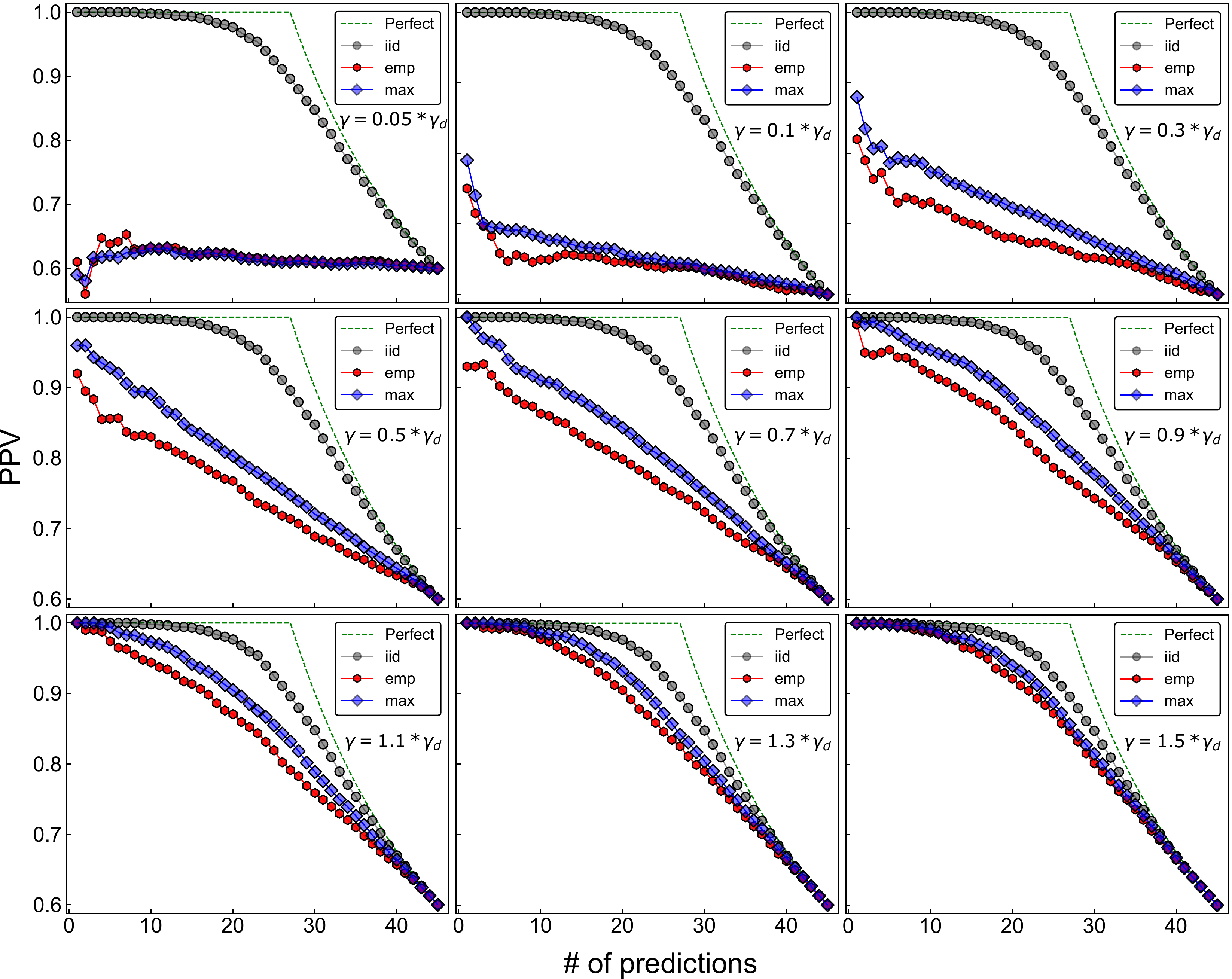}
	\caption{Quality of prediction of interactions for  different values of $\gamma$ and system size $L=10$. Interactions are defined as non-zero elements of the coupling matrix. In the $L=10$ case, there are $45$ possible interactions. Predictions are made by taking the largest elements (in absolute terms) of the inferred coupling matrix. The PPV is the fraction of correctly predicted contacts for a given number of predictions.}
	\label{fig:PPV_L10}
\end{figure}

\end{document}